\pdfoutput=1
\documentclass[12pt]{article}
\usepackage{amsmath}

\usepackage{amssymb}
\usepackage{graphicx}
\usepackage{isotope}
\usepackage{color}
\usepackage{float}
\newcommand{\be}{\begin{equation}}
\newcommand{\ee}{\end{equation}}
\newcommand{\bea}{\begin{eqnarray}}
\newcommand{\eea}{\end{eqnarray}}

\catcode`@=12

\begin{document}
\thispagestyle{empty}
\begin{center}
{\Large\bf
{Cooling of Dark-Matter Admixed Neutron Stars with  density-dependent Equation of State}}\\
\vspace{1cm}
{{\bf Sajad A. Bhat} \footnote{email: sajad.bhat@saha.ac.in} and
{\bf Avik Paul} \footnote{email: avik.paul@saha.ac.in}}\\
\vspace{0.25cm}
{\normalsize \it Astroparticle Physics and Cosmology Division,}\\
{\normalsize \it Saha Institute of Nuclear Physics, HBNI} \\
{\normalsize \it 1/AF Bidhannagar, Kolkata 700064, India}\\
\vspace{1cm}
\end{center}
\begin{abstract}
We propose a dark-matter (DM) admixed density-dependent equation of state where the fermionic DM interacts with the nucleons via Higgs portal. Presence of DM can hardly
influence the particle distribution inside neutron star (NS) but can significantly affect the structure as well as equation of state (EOS) of NS. 
Introduction of DM inside NS softens the equation of state.
We explored the effect of variation of DM mass and  DM Fermi momentum on the NS EOS. Moreover, DM-Higgs coupling is 
constrained using dark matter direct detection experiments. Then, we studied cooling of normal NSs using APR and DD2 EOSs and DM admixed NSs using 
dark-matter modified DD2 with varying DM mass and Fermi momentum. We have done our analysis by considering different NS masses. Also DM mass and DM Fermi momentum are varied for fixed 
NS mass and DM-Higgs coupling. We calculated the variations of luminosity and temperature of
NS  with time for all EOSs considered in our work and then compared our calculations with the observed astronomical cooling data of pulsars namely Cas A, RX
J0822-43, 1E 1207-52, RX J0002+62, XMMU J17328, PSR B1706-44, Vela, PSR B2334+61,
PSR B0656+14, Geminga, PSR B1055-52 and RX J0720.4-3125. It is found that APR EOS 
agrees well with the pulsar data for lighter and medium mass NSs but cooling is very fast for heavier NS. For DM admixed DD2 EOS, it is found that for all considered NS masses, all chosen DM masses and Fermi momenta agree well with the observational data of PSR B0656+14, Geminga, Vela, PSR B1706-44 and PSR
B2334+61. Cooling becomes faster as compared to normal NSs in case of increasing
DM mass and  Fermi momenta. It is infered from the calculations that if low mass super cold NSs are observed in future that may support the fact that heavier WIMP can be present inside neutron stars.  
\end{abstract}
\section{Introduction}
Neutron stars are excellent celestial laboratories for investigating the supradense nuclear matter which is otherwise inaccessible to terrestrial laboratories. 
The density inside neutron stars is several times the saturated nuclear density, hence exotic particles like hyperons \cite{Glenda:1985,{Glenda:1997}}, pion
or kaon condensate \cite{Rho:1979,{Kaplan:1986}}and quarks \cite{Collins:1975} are believed
to be present inside the core. Dark matter (DM) particles may also be captured and accumulated inside neutron stars \cite{Panotopoulos:2017idn}- \cite{Quddus:2019ghy}. Exotic particles soften the equation of state (EOS) and reduce the tidal deformability of the neutron star \cite{Bandyopadhyay:2018}. Exact nature of the matter is a challenging task and yet to be known. Any model suggested should not only describe the superdense matter
but also reproduce the properties of matter observed at saturation densities \cite{Walecka:1974qa,{Serot:1997xg}}. Recently, the unprecedented joint detection of  neutron star merger GW170817 by Advanced LIGO and 
Virgo observatories has put stronger constraints on the equation of state by constraining tidal deformability of NSs \cite{Abbott:2017a,{Abbott:2017}}. Using Shapiro delay measurements, a very massive neutron star
has been found in the form  of PSRJ0740+6620 with mass $ 2.17 ^{+0.11}_{-0.10}$ \cite{Cromartie:2019kug}. This can put stringent constraint on the equation of state.

Nowadays there are various cosmological and astrophysical indications for  the existence of dark matter in the Universe like rotation curves of 
spiral galaxies, large-scale structures of the Universe, anisotropies of cosmic microwave background radiation (CMBR), gravitational lensing etc. The detection of dark 
matter is attempted following three different ways i.e. direct detection, indirect detection and collider searches (LHC). However, till now no experimental signature of 
dark matter  has
been discovered.  Direct detection experiments put  upper bounds on the dark matter-nucleon elastic scattering cross-sections for different DM masses. In the literature,  many theoretical particle
dark matter models are  proposed to indirectly detect the dark matter and to explain the  existence of few unsolved phenomenological evidences such as gamma ray excesses observed 
by Fermi-LAT
gamma ray telescope \cite{Weniger:2012tx, {Bringmann:2012vr}}, positron excesses measured by PAMELA \cite{Adriani:2008zr}, AMS-02 \cite{AMS-02}, DAMPE \cite{Ambrosi:2017wek} experiments etc. Till now many particle candidates of dark matter are proposed like Weakly
Interacting Massive Particles (WIMPs) \cite{Jungman:1995df}-\cite{Biswas:2013nn}, Axions \cite{Peccei:2006as,{Weinberg:1977ma}}, Febbly Interating Massive Particles (FIMPs) \cite{Yaguna:2011qn,{Molinaro:2014lfa}}, Fuzzy dark matter \cite{Lidz:2018fqo,{{Amorisco:2018dcn}}}, neutralino \cite{Jungman:1995df}, 
Kaluza Klein dark matter \cite{Bergstrom:2006ny} etc. In this work our proposed particle candidate of dark matter is WIMP. In the early Universe, WIMPs are produced thermally and initially they 
are at thermal equilibrium but when the temperature drops below the WIMPs mass they are decoupled at a particular temperature ($\sim \dfrac{M_{\chi}}{20}$) called freeze-out
temperature. After decoupling, WIMP would possibly be a relic particle and may constitute a particle candidate of cold dark matter (CDM). WIMPs can cluster with stars gravitationally
and also form a 
background density in the universe.

Several studies have indicated that
neutron stars being highly compact objets can capture more dark matter particles during the formation stage in the 
supernova explosion
as compared to
the non-compact objects \cite{Fuller:2014rza}. Recently, it is shown that the admixture of DM inside NSs softens the equation of state and hence tidal deformability is reduced \cite{Das:2018frc}.
It has been proven in \cite{Brayeur:2011yw} that the DM capture could be highly improved if it happens in binary pulsars. Since the DM present inside 
DM admixed NSs can possibly change the global properties of neutron stars, this open another indirect window to study DM apart from the other numerous ways.
The structures of DM admixed NSs have been studied recently. It is shown that mass and radius of NSs can be remarkably affected by mirror DM \cite{Sandin:2008db}. It has been shown that fermionic DM
could soften the equation of state and hence reduce the maximum mass supported by the NS \cite{Das:2018frc}. This effect is sensitive to the mass of DM particle and 
the self-interaction within the dark matter. Since the normal matter and DM are believed to interact gravitationally, presence of DM can hardly
influence the the particle distribution inside NS but can significantly affect the structure as well as EOS of NS.

Cooling of neutron stars have been well studied by several authors \cite{Page:2009fu}-\cite{Potekhin:2017ufy}. Some study has been done on the effect of DM on cooling of NSs \cite{Ding:2019}-\cite{Hamaguchi:2019oev}. It has been found that the heating due to dark matter annihilation can affect the temperature of the stars older than $10^7$ years and consequently flattening out
the temperature at $10^4$ K for the neutron stars \cite{Kouvaris:2007ay}. Moreover, recently it has been found that slowdown in the pulsar rotation can drive the NS matter out of beta equilibrium and the resultant imbalance in chemical potentials can induce late-time heating, named as rotochemical heating which can heat a NS up to $10^6$ K for $t=10^6-10^7$ years \cite{Hamaguchi:2019oev}. In Ref. \cite{Ding:2019}, the authors have studied the cooling of DM admixed NS with dark matter mass ranging from 0.1 GeV to 1.3 GeV. In the present work, we have considered low as well as high dark matter masses (upto 500 GeV) and also varied the dark matter Fermi momenta for the cooling calculations. For these calculations, we have considered dark-matter modified density-dependent (DD2) EOS \cite{Banik:2014,{Typel:2010}} and the results are compared with the observational data. Also, our work is different from the above mentioned works \cite{Kouvaris:2007ay, {Hamaguchi:2019oev}} because we don't consider heating due to WIMP annihilation owing to the very small annihilation cross-section. We consider indirect effect on cooling of NS stars due to change in the neutron star structure in presence of dark matter. With the introduction of dark matter, cooling properties 
can change
significantly as compared to the normal NSs mainly because of changes in neutrino emissivity, neutrino luminosity and heat capacity. For given mass, neutron emissivity will be
different due to significant change in stellar structures and consequently, neutrino luminosity will also be different. Heat capacity related to EOS will be different for normal NSs 
and DM admixed NSs because of softening of the EOS in case of the latter. Thus, normal NSs can be distinguished from DM admixed NSs using astronomical observation data related to
surface temperature and age   of 
pulsars. We have considered the complete set of cooling data of both young and cool and old and warm neutron stars namely Cas A, RX J0822-43, 1E 1207-52, RX J0002+62, XMMU J17328, PSR B1706-44, Vela, PSR B2334+61, PSR B0656+14, Geminga, PSR B1055-52 and RX J0720.4-3125. We adopted the temperature data sets of the above mentioned pulsars from the Refs. \cite{Grigorian:2018bvg} - \cite{Yanagi:2019vrr}
and the luminosity data from the Ref. \cite{Vigano:2013lea} (Table 1 and 3).
As representatives for late time-cooling,  a group of above mentioned three NSs (PSR B0656+14, Geminga and PSR B1055-52) is chosen  forming a class of nearby objects that allows spectral fits to their X-ray emission (\cite{DeLuca:2004ck}- \cite{Paul:2018msp} and references therein). We studied NS cooling of both normal NSs using DD2 EOS \cite{Banik:2014,{Typel:2010}}
and Akmal-Pandharipande-Ravenhall 
(APR) EOS \cite{Akmal:1998cf} and
DM admixed NSs using
DD2 EOS modified with DM sector. It is important to mention here that although DD2 
is marginally allowed by the tidal deformability constraint obtained from the analysis of GW170817 with Phenom PNRT model \cite{Bhat:2018erd}, DM admixed DD2 will be softened and
might be considerably allowed by the GW170817 constraints. Earlier DM admixed NSs are studied by some groups \cite{Panotopoulos:2017idn}- \cite{Quddus:2019ghy} where they adopted $\sigma$-$\omega$-$\rho$ model 
but our approach differs from theirs in the sense that meson-nucleon couplings are density-dependent in our model which gives rise to an extra term called rearrangement
term \cite{Typel:2010,{Banik:2002}} in the nucleon chemical potential. 

This paper is organised as follows. In section 2, we describe baryonic EOS model and DM admixed baryonic EOS model. We constrain dark matter-Higgs coupling parameter from the direct
detection experiments as discussed in section 3. In section 4, we discuss the cooling mechanism of neutron star. Furthermore, the results and calculations are presented in section 5.
Finally, we conclude our work in section 6.        

\section{Equation of State Model}
In this section, we utilize density-dependent relativistic hadron field theory for describing strongly interacting superdense nuclear matter inside neutron stars. 
Nucleon-nucleon strong interaction is mediated by the exchanges of scalar $\sigma$ meson, responsible for strong attractive force, 
vector $\omega$, responsible for strong repulsive force and $\rho$ meson, responsible for symmetry energy. The Lagrangian density \cite{Banik:2014,{Typel:2010}} is given by
\begin{eqnarray}
\label{eq:1}
{\cal L}_B &=& \sum_B \bar\psi_{B}\left(i\gamma_\mu{\partial^\mu} - m_B
+ g_{\sigma B} \sigma - g_{\omega B} \gamma_\mu \omega^\mu \right. \nonumber\\
&& \left. - g_{\rho B} 
\gamma_\mu{\mbox{\boldmath $\tau$}}_B \cdot 
{\mbox{\boldmath $\rho$}}^\mu  \right)\psi_B\nonumber\\
&& + \frac{1}{2}\left( \partial_\mu \sigma\partial^\mu \sigma
- m_\sigma^2 \sigma^2\right)
-\frac{1}{4} \omega_{\mu\nu}\omega^{\mu\nu}\nonumber\\
&&+\frac{1}{2}m_\omega^2 \omega_\mu \omega^\mu \nonumber\\
&&- \frac{1}{4}{\mbox {\boldmath $\rho$}}_{\mu\nu} \cdot
{\mbox {\boldmath $\rho$}}^{\mu\nu}
+ \frac{1}{2}m_\rho^2 {\mbox {\boldmath $\rho$}}_\mu \cdot
{\mbox {\boldmath $\rho$}}^\mu,
\label{had}
\end{eqnarray}
where $\psi_B$ denotes the nucleon fields, ${\mbox{\boldmath 
$\tau_{B}$}}$ is the isospin operator and $g$s represent the  density-dependent 
meson-baryon couplings. These couplings are determined by following the prescription adopted by Typel et al. \cite{Typel:2010,{Typel:2005}}.
The functional dependence of the couplings on density was introduced for the first time in \cite{Typel:1999} as described here 
 \begin{eqnarray} \label{eq:2}
  g_{\alpha B}(\rho_b) = g_{\alpha B}(\rho_0)f_{\alpha}(x),
 \end{eqnarray}
where {$\rho_b$} is the total baryon density, $x = \rho_b/\rho_0$ and $f(x) = a_{\alpha} \frac{1+b_{\alpha}(x+d_{\alpha})}{1+c_{\alpha}(x+d_{\alpha})} $
for $\alpha = \omega$, $\sigma$. In order to reduce parameters, the functions are constrained as  $f_{\sigma}(1)=f_{\omega}(1)=1$, ${f}^{\prime}_{\sigma}(0)={f}^{\prime}_{\omega}(0)=0$,
$f^{\prime\prime}_{\sigma}(1)=f^{\prime\prime}_{\omega}(1)$. Exponential density dependence i.e. $ f(x) = exp[-a_{\alpha}(x-1)]$ \cite{Typel:1999} is considered for the isovector meson {\boldmath $\rho_{\mu}$}
because $g_{\rho B}$ decreases at higher densities. Finite nuclear properties are fitted to determine the saturation density, the mass of $\sigma$ meson, 
the couplings $g_{\alpha B}(\rho_0)$ and the coefficients $a_{\alpha}$, $b_{\alpha}$, $c_{\alpha}$ and $d_{\alpha}$ \cite{Typel:2010,{Typel:2005}}. The fit provides the saturation density $\rho_0 = 0.149065 fm^{-3}$, 
binding energy per nucleon as -16.02 MeV and compressibility factor $K= 242.7$ MeV. The nucleon mass $M_n$ is considered to be $939$ MeV through out our work.

Leptons are treated as non-interacting particles and  described by the free Lagrangian density
\begin{eqnarray} \label{eq:3}
{\cal L}_l = \sum_l \bar\psi_{l}\left(i\gamma_\mu{\partial^\mu} - m_l\right) \psi_{l},
\end{eqnarray}
where $l = {e^{-},\mu^{-}}$ and $m_l = m_e$ , $m_{\mu}$. The energy and pressure due to leptons will be explicitly mentioned in Section 2.1.  

\begin{table}[H]
\centering
\begin{tabular}{|l|c|c|c|c|c|c|c|c|r|}
\hline
meson $\alpha$ &$m_{\alpha}$&$g_{\alpha B}(\rho_0)$&$a_{\alpha}$&$b_{\alpha}$&$c_{\alpha}$&$d_{\alpha}$\\
 &in MeV&&&&&\\
 \hline
$\omega$&783.0 &13.342362&1.369718&0.496475&0.817753&0.638452\\
$\sigma$&546.212459&10.686681&1.357630&0.634442&1.005358&0.575810\\
$\rho$&763.0&3.626940&0.518903&&&\\
\hline
\end{tabular}
\caption{ Meson masses and parameters of meson-nucleon couplings in DD2 EOS.}\label{t1}
\end{table}
 
\subsection{Effect of Dark Matter on Equation of State}
A uniformly distributed fermionic dark matter (WIMP) is considered inside neutron star. Dark matter interacts with  Higgs field $h$ with coupling strength $y$. DM-Higgs coupling $y$
is explicitly discussed in Section 3. Three different WIMP masses ($M_{\chi} = 50$ GeV, 200 GeV, 500 GeV) are
considered in  our calculations. Higgs field $h$ interacts with the nucleons via effective Yukawa coupling $fM_n/v$, 
where $f$ denotes the nucleon-Higgs form factor and is estimated to be approximately 0.3 \cite{cline:2015} and $v=246.22$ GeV denotes Higgs vacuum expectation value (VEV). In the Higgs potential, terms higher than quadratic 
are dropped because they are negligible in the mean field approximation (MFA). Hence the dark sector and its interaction with nucleons and Higgs field is described by the 
Lagrangian density 
\begin{eqnarray}
\label{eq:4}
{\cal L }_{DM} &=&  \bar\chi (i\gamma_\mu{\partial^\mu} - M_{\chi} + yh) \chi + \frac{1}{2} \partial_\mu h \partial^\mu h
- \frac{1}{2} M^{2}_{h} h^{2} + f\frac{M_n}{v} \bar\psi h \psi.
\end{eqnarray}
Here we consider the assumption that the average dark matter number density inside neutron star is $10^3$ times smaller than saturated nuclear matter number density \cite{Panotopoulos:2017idn, {Li:2012qf}} and the Fermi momentum of dark matter is constant \cite{Panotopoulos:2017idn} through out the neutron star.  With these assumptions, the fractional mass of dark matter inside neutron star for $M_{\chi}=200$ GeV can be expressed as \\

~~~~~~~~~~~~~~~~~~~~~~~~~~~~~~~~~~~~~~~~~~~~~~$\dfrac{M_{\chi}}{M_{NS}} \approx \dfrac{1}{6}$. \\
Given $\rho_0 = 0.149065 fm^{-3}$, dark matter number density is $\rho_{DM} \sim 10^{-3}\rho_0 \sim 0.15 \times 10^{-3} fm^{-3}$. Number density of dark matter is related to Fermi momentum via 
~$\rho_{DM} = \frac{\left(k^{DM}_{F}\right)^3}{3\pi^2}$ which gives $k^{DM}_{F} \sim 0.033$ GeV. We vary $k^{DM}_{F}$ in our calculations from 0.01 GeV to 0.06 GeV and dark matter densties 
$\rho_{DM}$ will also vary accordingly. 
Equations of motion for nucleon doublet  \begin{align}
    \psi &= \begin{bmatrix}
           \psi_{p} \\
           \psi_{n}  \nonumber 
            \end{bmatrix},    
  \end{align}  
scalar meson ($\sigma$), vector meson ($\omega^{\mu}$) and isovector meson ({\boldmath $\rho ^\mu$}), DM particle ($\chi$) and 
Higgs boson $h$ can be derived from Eq. (\ref{eq:1}) and Eq. (\ref{eq:4}) as 

\begin{equation}
\label{eq:5}
 [\gamma^{\mu}(i \partial_\mu - \Sigma_B) - (M_n - g_{\sigma B} \sigma -\frac{f M_n}{v} h )] \psi_B = 0,  \nonumber\\
\end{equation}
\begin{eqnarray} \label{eq:6}
  \partial_{\mu} \partial^{\mu} \sigma + m^{2}_{\sigma} \sigma &=& g_{\sigma B} \bar\psi_B \psi_B , \nonumber \\
 \partial_{\mu} \omega^{\mu\nu} + m^{2}_{\omega} \omega^{\nu} &=& g_{\omega B} \bar\psi_B \gamma^{\nu} \psi_B, \nonumber \\
 \partial_{\mu} {\mbox {\boldmath $ \rho $ }}^{\mu \nu} + m^{2}_{\rho} {\mbox {\boldmath $\rho$ } }^{\nu} &=& g_{\rho B} \bar\psi_B \gamma^{\nu} {\mbox {\boldmath $\tau_B$ }} \psi_B, \nonumber \\
 (i\gamma_\mu{\partial^\mu} - M_{\chi} + yh) \chi &=& 0, \nonumber \\
  \partial_{\mu} h \partial^\mu h+  M^{2}_{h} h^{2} &=& y\bar\chi \chi + f\frac{M_n}{v} \bar\psi_B \psi_B,
  \end{eqnarray}
where masses of DM paticle and Higgs particle  are denoted  by $M_\chi$ and $M_h=125.09$ GeV respectively. $\Sigma_B = \Sigma^{0}_{B} + \Sigma^{r}_{B}$ is the vector self energy
in which the first term consists of the usual non-vanishing components of vector mesons i.e. $ \Sigma^{0}_{B} = g_{\omega B} \omega_{0} - g_{\rho B} \tau_{3B}{\bf \rho_{03}}$
 and the second term is the rearrangement term 
 i.e. $\Sigma^{r}_{B} = \sum_B [- g^{\prime}_{\sigma B} \sigma \rho^{s}_{B} +g^{\prime}_{\omega B} \omega_0 \rho_{B} +g^{\prime}_{\rho B} \tau_{3B} {\bf \rho_{03}} \rho_{B}] $ 
 which appears because
 of density-dependence of meson-nucleon couplings \cite{Banik:2002}. Here $g^{\prime}_{\alpha B}= \frac{\partial g_{\alpha B}}{\partial \rho_B}$ 
 where $\alpha = \sigma$, $\omega$, $\rho$ and $\tau_{3B}$ is the isospin projection of 
 $B = n$, p. Due to density dependence of nucleon-meson couplings, chemical potential of the nucleons takes the
 form
 
 ~~~~~~~~~~~~~~~~~~~~~~~~$\mu_B = \sqrt{k^{2}_{B} + M^{*2}_n} + \Sigma^{0}_{B} + \Sigma^{r}_{B} $.

In the mean-field approximation (MFA), fields are replaced by their expectation values and above equations are
simplified as 
\begin{eqnarray} \label{eq:8}
 \sigma &=& \frac{1}{m^{2}_{\sigma}} (g_{\sigma B}\langle \bar\psi_B \psi_B \rangle ), \nonumber \\
 \omega_0 &=& \frac{g_{\omega B}}{m^{2}_{\omega}} \langle \psi^{\dagger}_B  \psi_B \rangle = \frac{g_{\omega B}}{m^{2}_{\omega}} (\rho_p + \rho_n), \nonumber \\
 h_0 &=& \frac{y \langle \bar\chi \chi \rangle + f\frac{M_n}{v} \langle \bar\psi_B \psi_B \rangle } {M^{2}_h}, \nonumber \\
 \rho_{03} = \frac{g_{\rho B}}{m^{2}_{\rho}} \langle \psi^\dagger_B \tau_{3B} \psi_B \rangle &=& \frac{g_{\rho B}}{m^{2}_{\rho}} (\rho_p - \rho_n), \nonumber \\
 (i \gamma^{\mu} \partial_{\mu} - \Sigma_B - M^{*}_{n}) \psi_B &=& 0, \nonumber \\
 (i \gamma^{\mu} \partial_{\mu}  - M^{*}_{\chi}) \chi &=& 0.
\end{eqnarray}
The effective masses of nucleons and dark matter are respectively given as 
\begin{eqnarray} \label{eq:9}
M^{*}_{n}     &=&  M_n - g_{\sigma B} \sigma -\frac{f M_n}{v} h_0, \nonumber \\
M^{*}_{\chi} &=& M_{\chi} - yh_0.
\end{eqnarray}

The baryon density ($\rho$), scalar density ($\rho_s$) and dark matter density ($\rho^{DM}_s$) are
\begin{eqnarray} \label{eq:10}
 \rho = \langle \psi^\dagger \psi \rangle &=& \frac{\gamma}{(2\pi)^3} \int^{k_F}_{0} d^3k, \nonumber \\
 \rho_{s} = \langle \bar\psi \psi \rangle &=& \frac{\gamma}{(2\pi)^3} \int^{k_F}_{0} \frac{M^{*}_n}{\sqrt{k^{2} + M^{*2}_n}} d^3k,  \nonumber \\
  \rho^{DM}_{s} = \langle \bar\chi \chi \rangle &=& \frac{\gamma}{(2\pi)^3} \int^{k^{DM}_F}_{0} \frac{M^{*}_{\chi}}{\sqrt{k^{2} + M^{*2}_{\chi}}} d^3k,
 \end{eqnarray}
where $k_F$ and $k^{DM}_F$ are the Fermi momenta for  nucleonic matter and dark matter respectively and $\gamma = 2 $ is the spin degeneracy factor of 
nucleons. The masses of  mesons and meson-nucleon couplings at saturation density $\rho_0$ are given in Table 1 \cite{Typel:2010,{Typel:2005}}. In order to get the density dependent profile for $M^{*}_n$ and $M^{*}_{\chi}$, Eqs. (\ref{eq:8}) and (\ref{eq:10})
should be solved self consistently. The energy and pressure i.e. EOS are provided by expectation values of
energy-momentum tensor in the static case as $\epsilon = \langle T^{00} \rangle $ and $P = \frac{1}{3}\langle T^{ii}\rangle$. \\
The total energy density and pressure for the combined Lagrangian ${\cal L}_B + {\cal L}_{DM}$ are obtained as 
\begin{eqnarray} \label{eq:11}
 \epsilon = g_{\omega B} \omega_0 (\rho_p + \rho_n) + g_{\rho B} \rho_{03}(\rho_p - \rho_n) + \frac{1}{\pi^2} \int^{k^{p}_F}_{0} dk k^2 \sqrt{k^2 + M^{*2}_n} \nonumber \\
+ \frac{1}{\pi^2} \int^{k^{n}_F}_{0} dk k^2 \sqrt{k^2 + M^{*2}_n} + \frac{1}{\pi^2} \int^{k^{DM}_F}_{0} dk k^2 \sqrt{k^2 + M^{*2}_{\chi}} \nonumber \\
 + \frac{1}{2} m^{2}_{\sigma} \sigma^{2} - \frac{1}{2} m^{2}_{\omega} \omega^{2}_{0} - \frac{1}{2} m^{2}_{\rho} \rho^{2}_{03} + \frac{1}{2} M^{2}_{h} h^{2}_{0},
 \end{eqnarray}
\begin{eqnarray} \label{eq:12}
 P =  \frac{1}{3\pi^2} \int^{k^{p}_F}_{0} dk \frac{ k^4} {\sqrt{k^2 + M^{*2}_n}} + \frac{1}{3\pi^2} \int^{k^{n}_F}_{0} dk \frac{ k^4} {\sqrt{k^2 + M^{*2}_n}} \nonumber \\
+ \frac{1}{3\pi^2} \int^{k^{DM}_F}_{0} dk \frac{k^4} {\sqrt{k^2 + M^{*2}_{\chi}} } \nonumber \\
 - \frac{1}{2} m^{2}_{\sigma} \sigma^{2} + \frac{1}{2} m^{2}_{\omega} \omega^{2}_{0} + \frac{1}{2} m^{2}_{\rho} \rho^{2}_{03} - \frac{1}{2} M^{2}_{h} h^{2}_{0},
 \end{eqnarray}
 where $\rho_n$ and $\rho_p$ are the neutron and proton number densities and $k^{n}_F$ and $k^{p}_F$ are the corresponding Fermi momenta of neutron and proton, respectively. 
The nuclear matter inside the neutron star will be charge neutral and $\beta$-equilibrated. The conditions of charge neutrality and $\beta$-equilibrium are given as
\begin{eqnarray} \label{eq:13}
 \rho_p = \rho_e + \rho_{\mu},
\end{eqnarray}
and 
\begin{eqnarray} \label{eq:14}
 \mu_n &=& \mu_p +\mu_e, \nonumber \\
 \mu_e &=& \mu_{\mu},
\end{eqnarray}
respectively. Here, the chemical potentials  $\mu_e$ and $\mu_{\mu}$ are given as 
\begin{eqnarray} \label{eq:15}
 \mu_e &=& \sqrt{k^{2}_{e}+ m^{2}_{e}},  \nonumber \\
\mu_{\mu} &=& \sqrt{k^{2}_{\mu}+ m^{2}_{\mu}},
\end{eqnarray}
whereas the nucleon chemical potentials contain the rearrangement term also because of density-dependence of couplings as mentioned earlier.
The particle fractions of neutron, proton, electron and muon will be determined by the  self consistent solution of Eqs. (\ref{eq:13}) and (\ref{eq:14}) for a given baryon density. 
The energy density and pressure due to the non-interacting leptons are given as
\begin{eqnarray} \label{eq:16}
 \epsilon_l &=& \frac{1}{\pi^2} \sum_l \int^{k^{l}_F}_{0} dk k^2 \sqrt{k^2 + m^{2}_l},  \\
  P_l &=& \frac{1}{3\pi^2} \sum_l \int^{k^{l}_F}_{0} dk \frac{k^4} {\sqrt{k^2 + m^{2}_{l}} }.
 \end{eqnarray}
So the total energy density and pressure of the charge neutral $\beta$-equilibrated neutron star matter are 
\begin{eqnarray} \label{eq:17}
 \epsilon_{NM} &=& \epsilon_l + \epsilon, \\
 P_{NM} &=& P_l + P.
\end{eqnarray}

\begin{figure}[H]
\includegraphics[width=5.5cm,height=6.4cm]{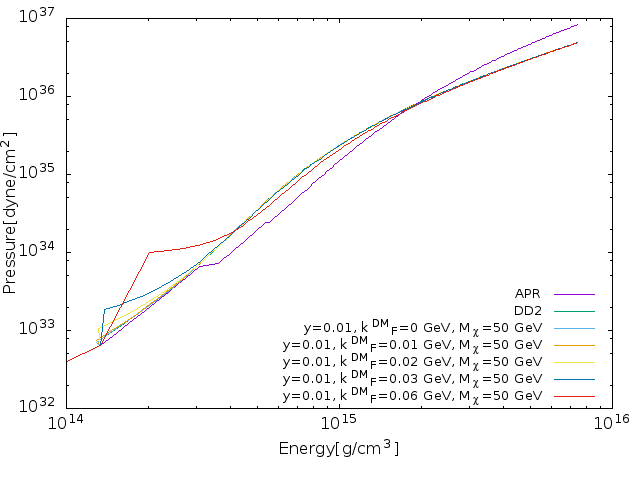}
\includegraphics[width=5.5cm,height=6.4cm]{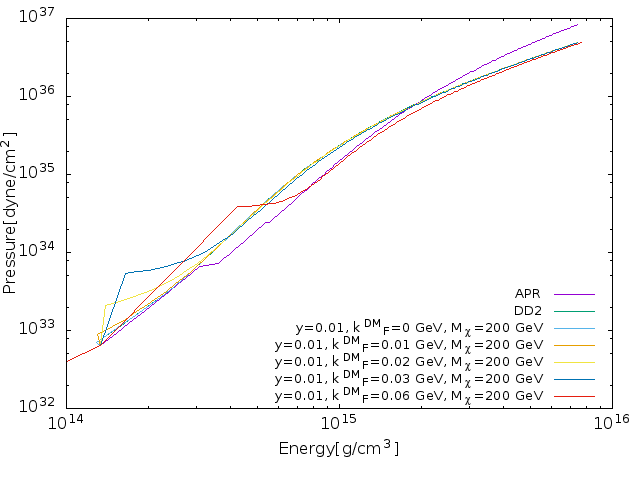}
\includegraphics[width=5.5cm,height=6.4cm]{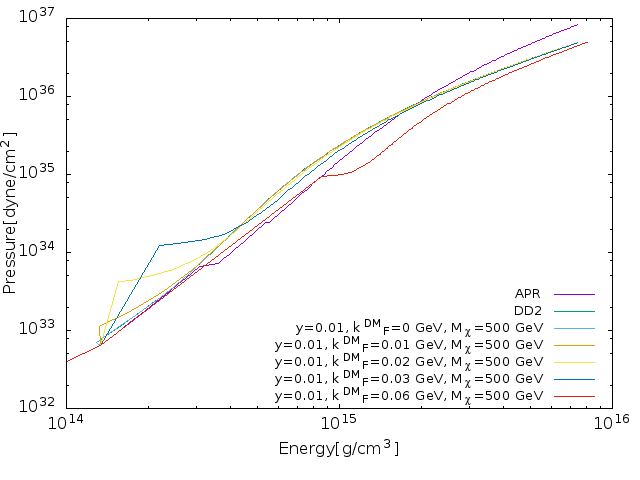}
\caption{Pressure vs energy plots for $M_\chi = 50$ GeV (Left panel), $M_\chi = 200$ GeV (middle panel), $M_\chi = 500$ GeV (right panel) with varying DM Fermi momenta in each panel.}
\label{fig:1}
\end{figure}

\begin{figure}[H]
\begin{center}
\includegraphics[width=5.5cm,height=6.5cm]{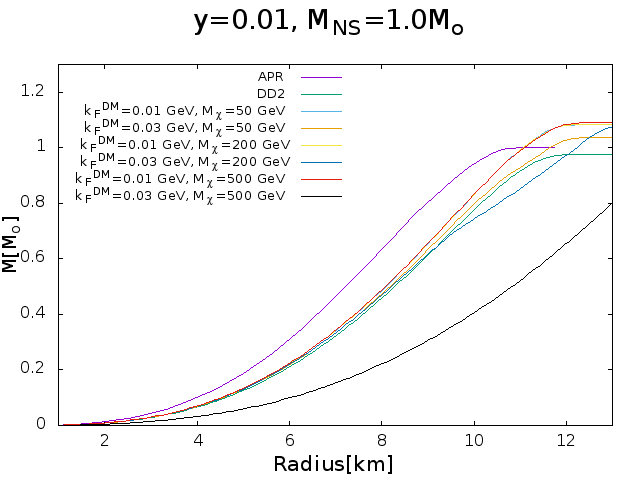}
\includegraphics[width=5.5cm,height=6.5cm]{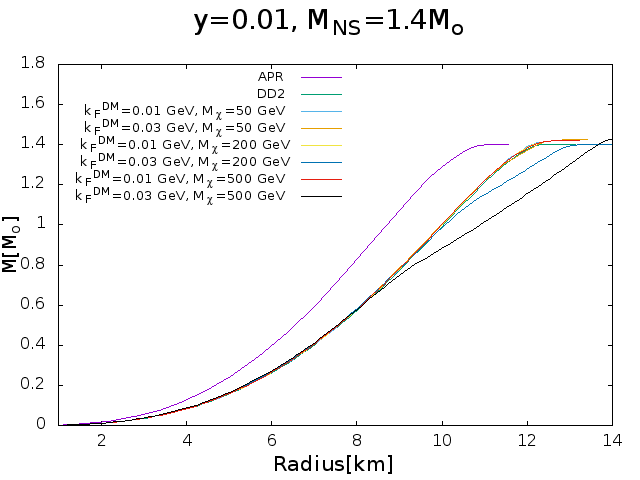}
\includegraphics[width=5.5cm,height=6.5cm]{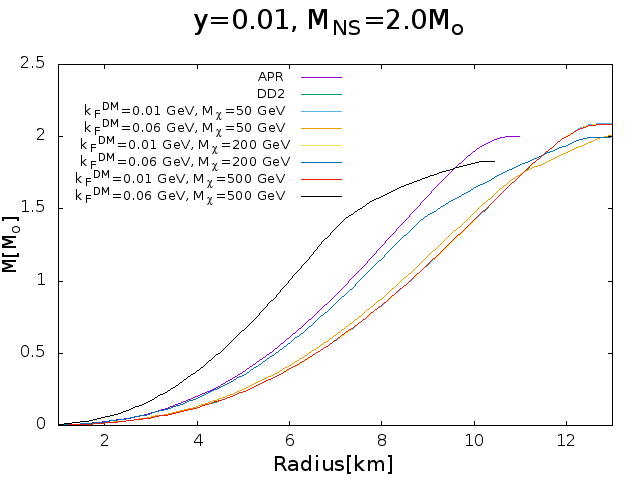}
\end{center}
\caption{Enclosed mass vs radius plots for $M_{NS} = 1.0 M_{\odot}$  (Left panel), $M_{NS} = 1.4 M_{\odot}$ (middle panel), $M_{NS} = 2.0 M_{\odot}$ (right panel) with varying DM mass and Fermi momentum in each panel.}
\label{fig:2}
\end{figure}

\begin{figure}[H]
\begin{center}
\includegraphics[width=5.5cm,height=6.5cm]{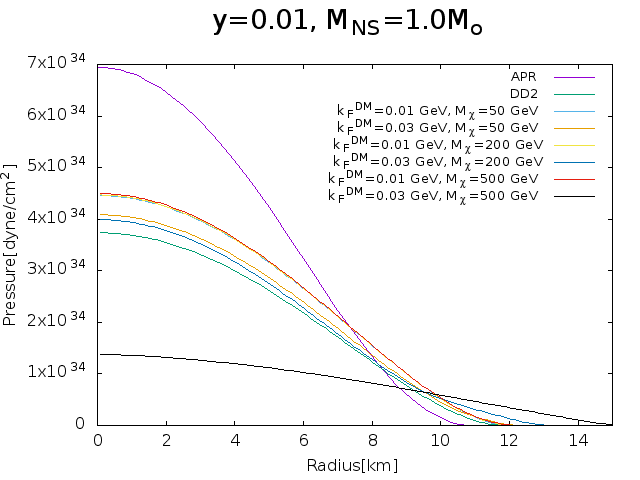}
\includegraphics[width=5.5cm,height=6.5cm]{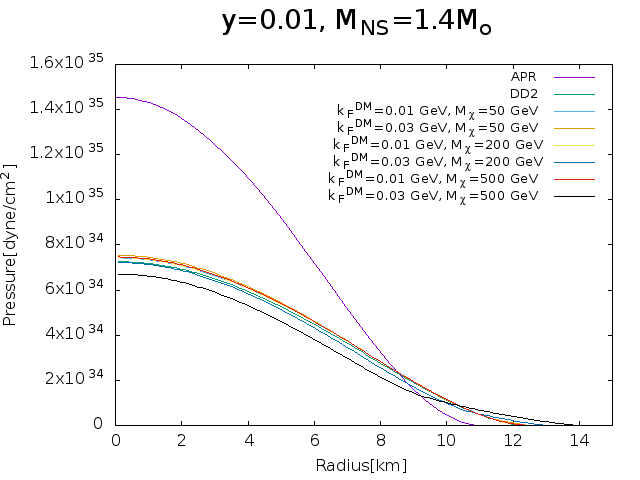}
\includegraphics[width=5.5cm,height=6.5cm]{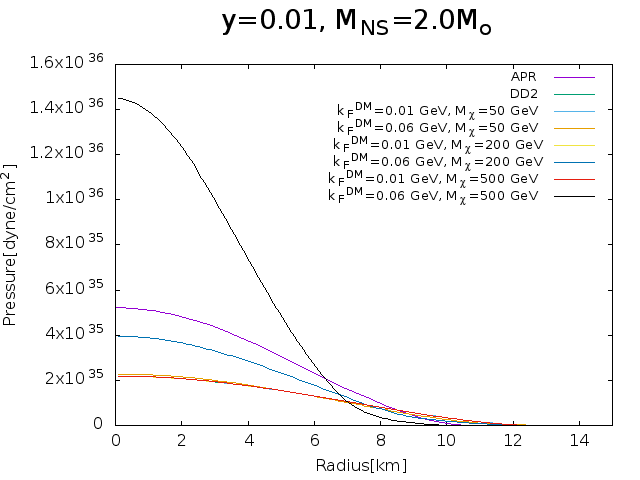}
\end{center}
\caption{Pressure vs radius plots for $M_{NS} = 1.0 M_{\odot}$  (Left panel), $M_{NS} = 1.4 M_{\odot}$ (middle panel), $M_{NS} = 2.0 M_{\odot}$ (right panel) with varying DM mass and Fermi momentum in each panel. }
\label{fig:4}
\end{figure}

For all the EOSs considered in our work, we solve numerically Tolman-Oppenheimer-Volkoff (TOV) \cite{Oppenheimer:1939} equations of hydrosatic equilibrium to generate the mass-radius and pressure-radius profiles
as shown in Figures \ref{fig:2}-\ref{fig:4}.
In Figure \ref{fig:1}, we present EOSs for different DM masses and Fermi momenta along with APR and DD2 EOSs. It is important to mention that all of these EOSs satisfy the causality condition i.e. $c_s^2 < 1$. It is evident that for a fixed DM-Higgs
coupling $y$ and fixed DM mass, EOS becomes softer for higher values of DM Fermi momentum and is softest for $k^{DM}_F = 0.06$ GeV for lower to moderate values of density and APR is stiffest
for higher values of density. Moreover, it is inferred from the comparison of three panels of Figure \ref{fig:1} that for fixed values of $y$ and $k^{DM}_{F}$, higher values of DM masses leads to 
softer EOS. It is important to mention here that for the higher DM mass $M_{\chi} = 500$GeV, the EOS corresponding to $k^{DM}_F = 0.06$ GeV becomes softest among all
 the cases of  DM masses. This sudden softening of EOS for $k^{DM}_F = 0.06$ GeV at $M_{\chi} = 500$GeV might be due to dominance of dark matter over baryonic matter at such extreme parameters of
DM. Nevertheless all the neutron star configurations for the dark matter densities considered in this work are stable and will not undergo black hole formation \cite{Li:2012qf}. In Figure \ref{fig:2}, mass-radius profile is plotted for NS masses 
$1M_{\odot}$, $1.4 M_{\odot}$ and $2M_{\odot}$. These plots can be explained the same way as in case of Figure 1. Here also mass-radius profile for $k^{DM}_F = 0.06$ GeV 
at $M_{\chi} = 500$GeV follows a trend contrary to other combinations of $k^{DM}_F$ and $M_{\chi}$. In this case the majority of mass contribution is from DM and hence
leading to smaller radius than other cases due to enhanced gravitational contraction. Figure \ref{fig:4} shows the pressure-radius profile for different NS masses where it is evident that for fixed
DM mass and NS mass, higher values of $k^{DM}_F$ leads to lower pressure except for the case of $M_{\chi} = 500$ GeV and $M_{NS}= 2M_{\odot}$ where $k^{DM}_F = 0.06$ GeV leads to higher pressure in the inner region of 
the star. This is because the star becomes more centrally condensed at very high DM mass.


\section{Direct Detection}
Dark matter direct detection experiments do not show any signatures of collision events as of now. These experiments give an upper bound on the elastic scattering cross-section 
as a function of dark matter mass. In the present scenario, fermionic dark matter (WIMP) can undergo an elastic collision with the detector nucleus (quark level) 
by the Higgs exchange. Therefore, the effective Lagrangian contains the scalar operator $\bar{\chi }\chi\bar{q}q$ and can be written as
\begin{equation}\label{eq:20}
\mathcal{L}_{\text{eff}}=\alpha_{q}\bar{\chi }\chi\bar{q}q,
\end{equation}
where $q$ represent the valence quarks and $\alpha_{q}=yf(\frac{m_q}{v})\left(\dfrac{1}{m_h^2}\right)$. This scalar operator contributes to the spin independent (SI)
scattering cross-section for the fermionic dark matter candidate and can be expression as
\begin{equation}\label{eq:21}
\left(\sigma_{{\rm{SI}}}\right)=\dfrac{y^2 f^2 M_n^2}{4\pi} \dfrac{m_r^2}{v^2 M_h^4}.
\end{equation}
In Eq. (\ref{eq:21}), $m_r=\frac{M_n M_{\chi }}{M_n+M_{\chi }}$ is the reduced mass. We calculate the SI scattering cross-section using Eq. (\ref{eq:21}) and 
then constrain the parameter $``y"$ using the direct detection experiments in such a way that calculated scattering cross-section for different dark matter masses are below the
experimental bounds. In the present scenario, we use XENON-1T 
\cite{Aprile:2015uzo}, PandaX-II  \cite{Tan:2016zwf}, LUX \cite{Akerib:2016vxi} and DarkSide-50 \cite{Marini:2016haq} experimental bounds for 
constraining the parameter $``y"$. We checked that by varying the value of the parameter $``y"$ in the cooling calculations,  no significant differences are found and hence, we accordingly fixed $``y"$ to be 0.01 and calculate the corresponding scattering cross-section for three
chosen DM masses as tabulated in Table  2.

\begin{table}[H]
\centering
\begin{tabular}{|l|c|c|c|c|c|c|c|c|r|}
\hline
\hspace{2mm}$m_{\chi}$&$y$&$\sigma_{{\rm{SI}}}$\\
 in GeV&&$\rm{cm^2}$\\
 \hline
50&0.01&1.4115$\times 10^{-47}$\\
\hline
200&0.01&1.4514$\times 10^{-47}$\\ 
\hline
500&0.01&1.4596$\times 10^{-47}$\\
\hline
\end{tabular}
\caption{Calculated values of spin independent DM-nucleon scattering cross-section  for three chosen DM masses at fixed DM-Higgs coupling $y$.}\label{t2}
\end{table}

\section{Cooling Mechanism of Neutron Stars}

It is well known  that the surface temperature of the neutron star  decreases with time which is the direct indication of cooling. In order to calculate the thermal evolution of 
 the neutron star, one needs to solve 
 the the energy balance equation for the neutron star which can be expressed as \cite{Page:2005fq}
\begin{equation} \label{eq:22}
\frac{dE_{\rm th}}{dt}=C_{v}\frac{dT}{dt}=
-L_{\nu}(T)-L_{\gamma}(T_{e})+H(T)
\end{equation}
where $E_{\rm th}$ represents the thermal energy content of the star, 
$T$  and $T_{e}$ are the   internal and effective
temperatures of the star, respectively, $C_{v}$ is the  heat capacity of the
core and  $H$ is the source term which includes different 
``heating mechanisms"  important in the 
later stage   of  neutron star evolution.  
  In Eq. (\ref{eq:22}),  $L_{\nu}$ and $L_{\gamma}$  denote the neutrino and photon
luminosities, respectively. $H(T)$ is considered here to zero.
The photon luminosity is  calculated using the Stefan-Boltzmann law \cite{Nscool:2010dpl} as 
\begin{equation} \label{eq:23}
L_{\gamma}=S\hspace{1mm}T^{2+4\alpha}=4\pi\hspace{1mm}\sigma\hspace{1mm}R^{2}\hspace{1mm}T_{e}^{4}
\hspace{2mm}.
\end{equation}
This  relation is obtained using $\hspace{2mm} T_{e}\propto T^{0.5+\alpha} \hspace{1mm}(\alpha\ll 1)$,
where $R$ is the radius of the star and  $\sigma$ denotes the Stefan-Boltzmann constant. NSCool code \cite{Nscool:2010dp} is utilised in the present work 
for calculating the neutrino and photon  luminosities. Several  neutrino emitting  processes contribute in the cooling of neutron stars \cite{Paul:2018msp, {Page:2005fq},{{Yakovlev:2000jp}}}.
  Direct Urca processes and modified
Urca processes are the  two main neutrino emitting processes for the cooling. The direct Urca processes are 

\hspace{4cm} $n\rightarrow p+e^{-
}+\overline{\nu}_{e}$, $p+e^{-}\rightarrow n+\nu_{e}$. 

\hspace{-0.8cm} These are  possible in neutron stars only  if the proton fraction crosses a critical threshold.
The  two processes are fast and the luminosity varies with the temperature as $L_{\nu}^{fast}\propto
T^{6}_{9}$. Modified Urca process will become more dominant provided the proton fraction is below the threshold.
 The modified processes are 
 
 $n+n\rightarrow n+p+e^{-}+\overline{\nu}_{e}$, $n+p+e^{-}\rightarrow n+n+\nu_{e}$, $p+n\rightarrow p+p+e^{-}+\overline{\nu}_{e}$ and $p+p+e^{-}\rightarrow p+n+\nu_{e}$.
 
\hspace{-0.6cm}These  are slow processes and  luminosity varies with the temperature as
$L_{\nu}^{slow}\propto T^{8}_{9}$. Cooper pairing  of nucleons are the other set of neutrino  emiitting processes as

\hspace{4cm}$n+n\rightarrow [nn]+\nu +\overline{\nu}$, $p+p\rightarrow [pp]+\nu +\overline{\nu}$. 

\hspace{-0.6cm}These are medium processes where luminosity varies with temperature as $L_{\nu}^{medium}\propto
T^{7}_{9}$. 
There are several other neutrino emitting processes involved in the cooling as follows

\hspace{3cm}$e^{-}+e^{+}\rightarrow \nu +\overline{\nu}$ (electron-positron pair annihilation),

\hspace{3cm}$e^{-}\rightarrow e^{-}+\nu +\overline{\nu}$ (electron synchrotron),

\hspace{3cm}$\gamma+e^{-}\rightarrow e^{-}+\nu +\overline{\nu}$ (photoneutrino emission),

\hspace{3cm}$e^{-}+Z\rightarrow e^{-}+Z+\nu +\overline{\nu}$ (electron-nucleus bremsstrahlung),

\hspace{3cm}$n+n\rightarrow n+n+\nu +\overline{\nu}$ (neutron-neutron bremsstrahlung) and

\hspace{3cm}$n+p\rightarrow n+p+\nu +\overline{\nu}$ (neutron-nucleus bremsstrahlung).

\section{Calculations and Results}
\begin{figure}[H]
\includegraphics[width=5.5cm,height=6.4cm]{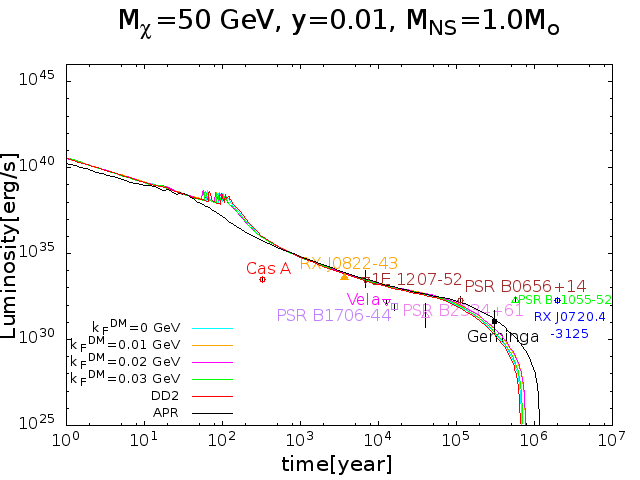}
\includegraphics[width=5.5cm,height=6.4cm]{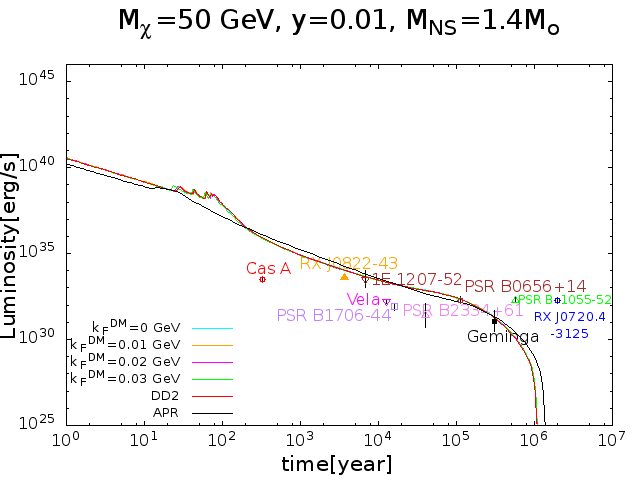}
\includegraphics[width=5.5cm,height=6.4cm]{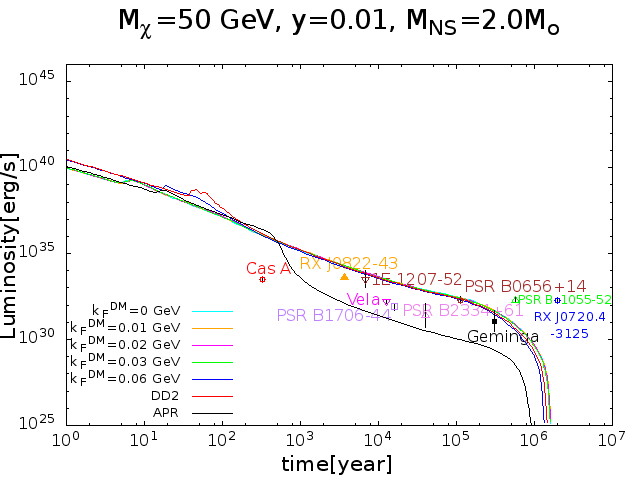}
\caption{Variation of luminosity with time three differently chosen NS masses $M_{NS} = 1.0 M_{\odot}$  (Left panel), $M_{NS} = 1.4 M_{\odot}$ (middle panel), $M_{NS} = 2.0 M_{\odot}$ (right panel) 
with varying $k^{DM}_F$ and fixed $M_\chi=50$ GeV in each panel. The theoretical calculations are compared with the observational data of pulsars namely Cas A, RX J0822-43, 1E 1207-52, PSR B1706-44, Vela, PSR B2334+61, PSR B0656+14, Geminga, PSR B1055-52 and RX J0720.4-3125 
shown by dots with error bars from left to right. }
\label{fig:5}
\end{figure}

\begin{figure}[H]
\includegraphics[width=5.5cm,height=7.5cm]{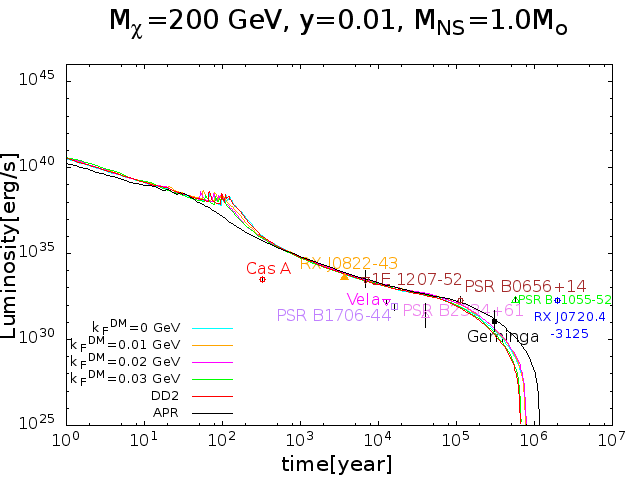}
\includegraphics[width=5.5cm,height=7.5cm]{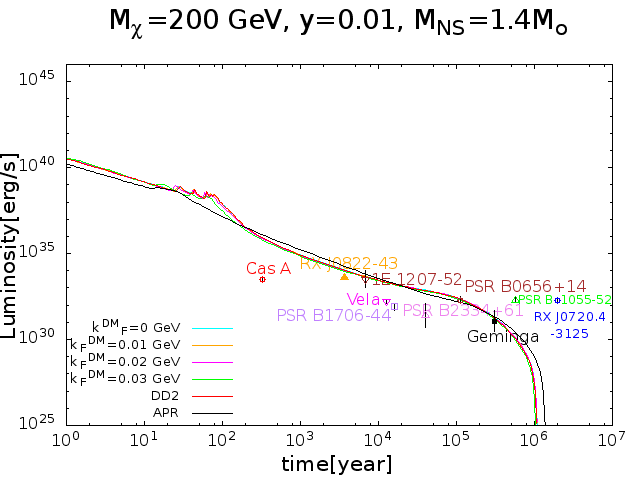}
\includegraphics[width=5.5cm,height=7.5cm]{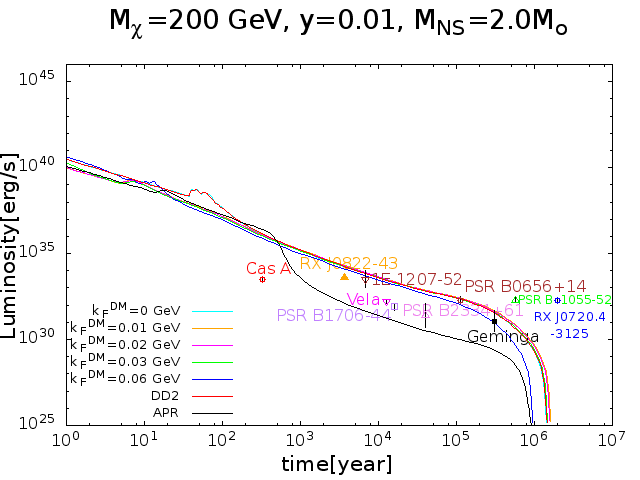}
\caption{Same as in Figure \ref{fig:5} but for $M_\chi=200$ GeV.}
\label{fig:6}
\end{figure}

\begin{figure}[H]
\includegraphics[width=5.5cm,height=7.5cm]{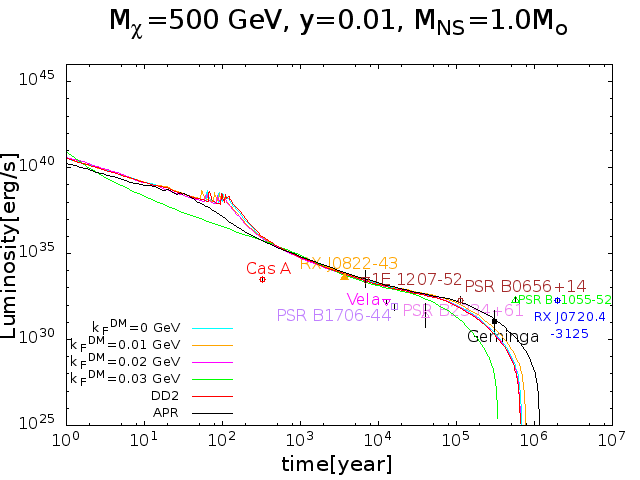}
\includegraphics[width=5.5cm,height=7.5cm]{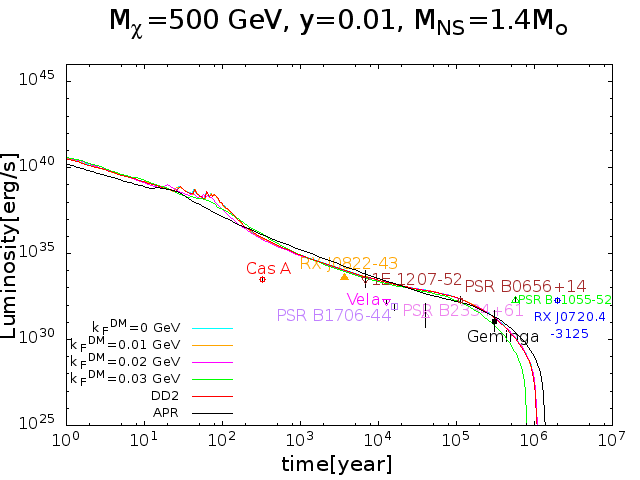}
\includegraphics[width=5.5cm,height=7.5cm]{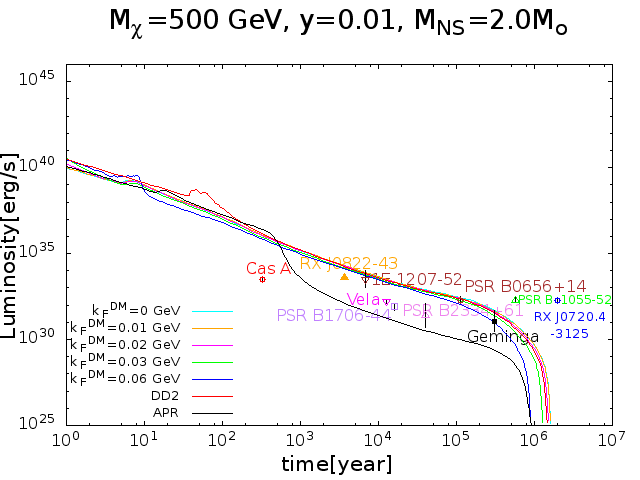}
\caption{Same as in Figure \ref{fig:5} but for $M_\chi=500$ GeV.}
\label{fig:7}
\end{figure}

\begin{figure}[H]
\includegraphics[width=5.5cm,height=6.5cm]{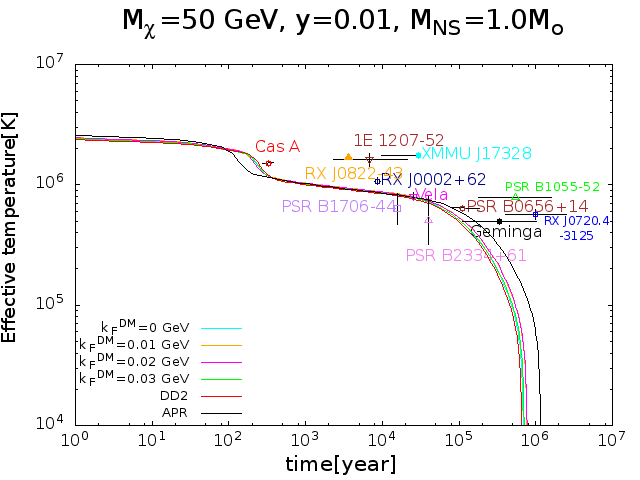}
\includegraphics[width=5.5cm,height=6.5cm]{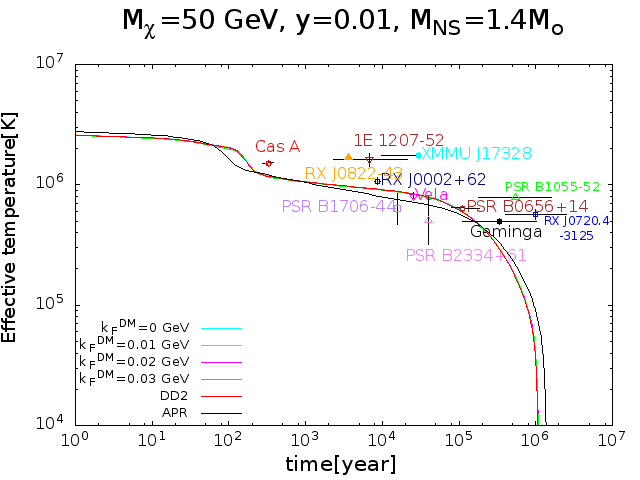}
\includegraphics[width=5.5cm,height=6.5cm]{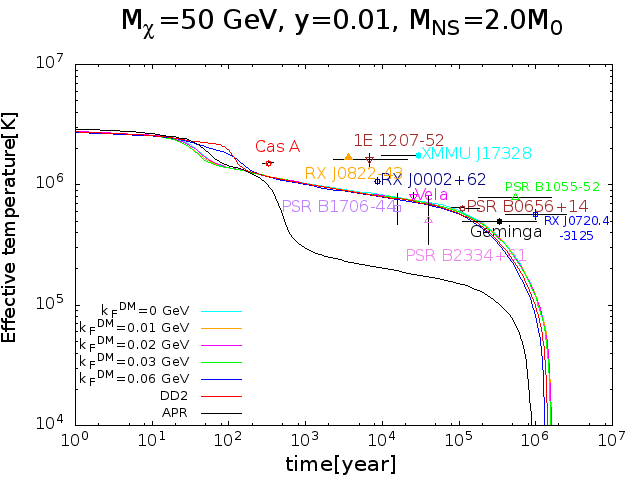}
\caption{Variation of effective temperature with time for three differently chosen NS masses $M_{NS} = 1.0 M_{\odot}$  (Left panel), $M_{NS} = 1.4 M_{\odot}$ (middle panel), $M_{NS} = 2.0 M_{\odot}$ (right panel) 
with varying $k^{DM}_F$ and fixed $M_\chi=50$ GeV in each panel. The theoretical calculations are compared with the observational data of pulsars namely Cas A, RX J0822-43, 1E 1207-52, RX J0002+62, XMMU J17328, PSR B1706-44, Vela, PSR B2334+61, PSR B0656+14, Geminga, PSR B1055-52 and RX J0720.4-3125 shown by dots with error bars from left to right.}
\label{fig:8}
\end{figure}

\begin{figure}[H]
\includegraphics[width=5.5cm,height=6.5cm]{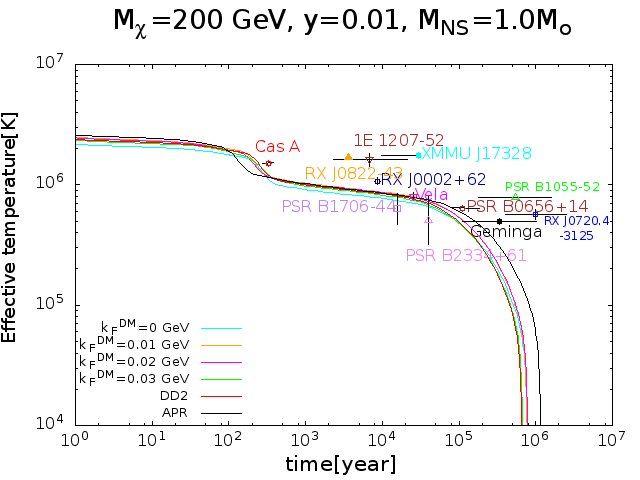}
\includegraphics[width=5.5cm,height=6.5cm]{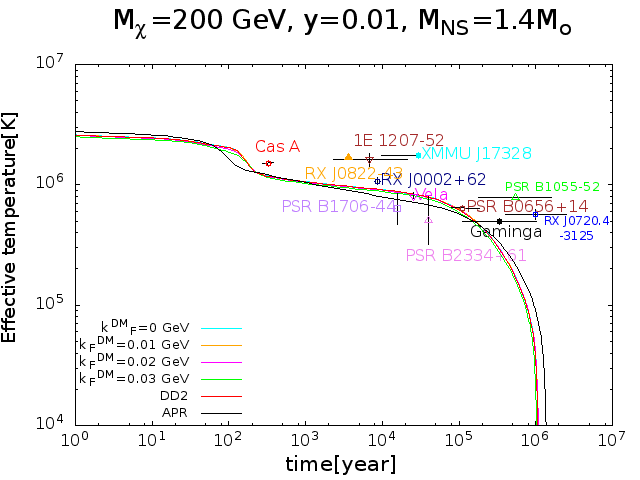}
\includegraphics[width=5.5cm,height=6.5cm]{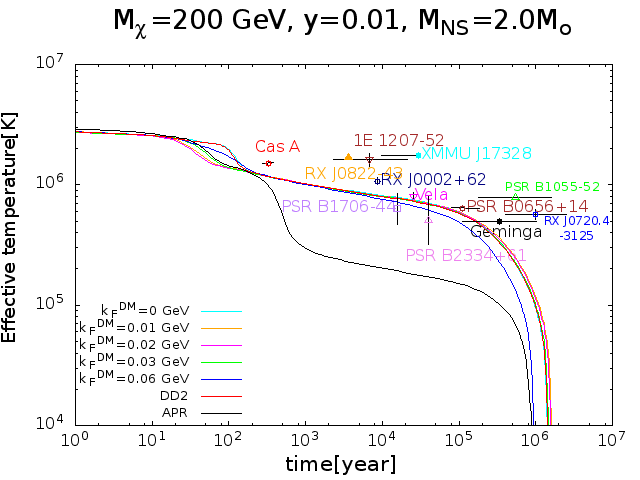}
\caption{Same as in Figure \ref{fig:8} but for $M_\chi=200$ GeV}
\label{fig:9}
\end{figure}

\begin{figure}[H]
\includegraphics[width=5.5cm,height=6.5cm]{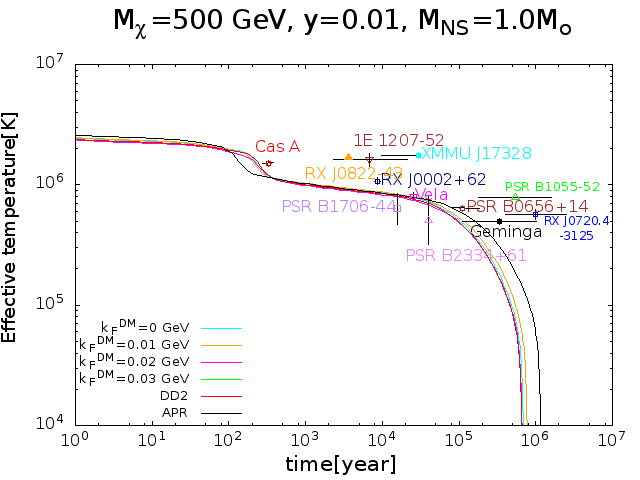}
\includegraphics[width=5.5cm,height=6.5cm]{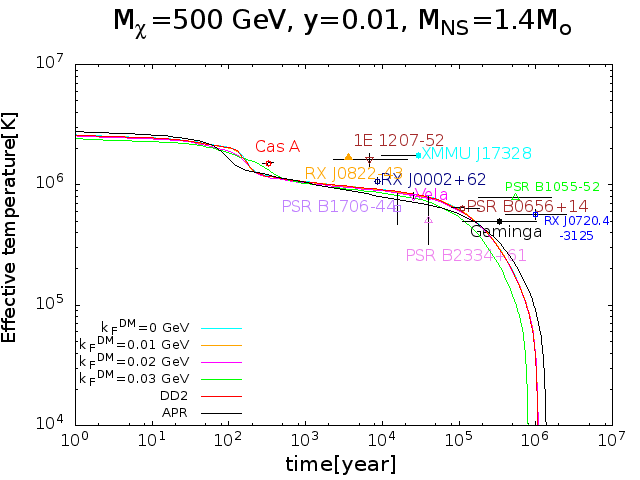}
\includegraphics[width=5.5cm,height=6.5cm]{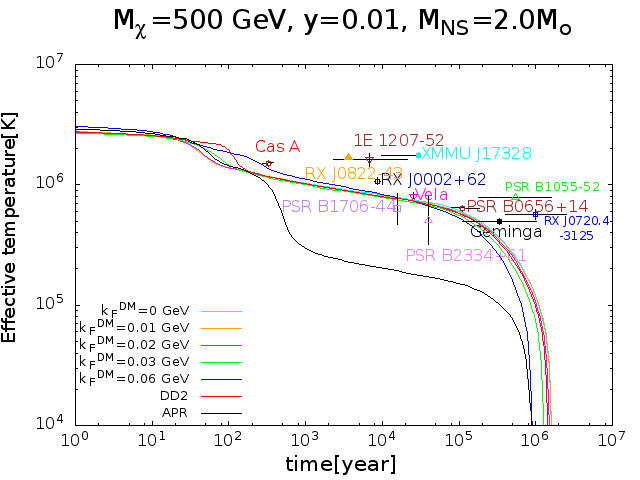}
\caption{Same as in Figure \ref{fig:8} but for $M_\chi=500$ GeV}
\label{fig:10}
\end{figure}

\begin{figure}[H]
\includegraphics[width=5.5cm,height=6.5cm]{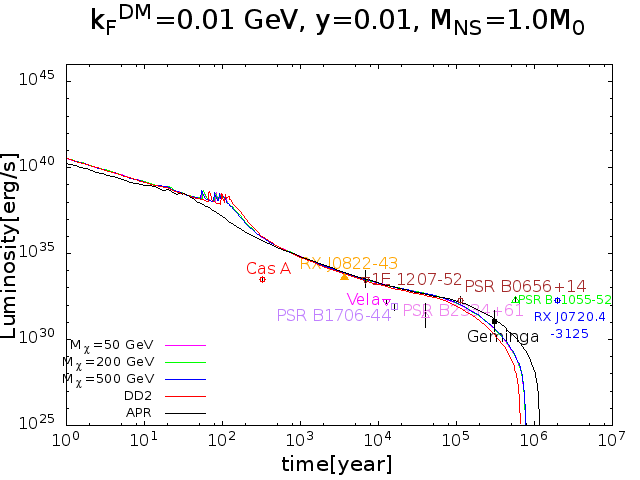}
\includegraphics[width=5.5cm,height=6.5cm]{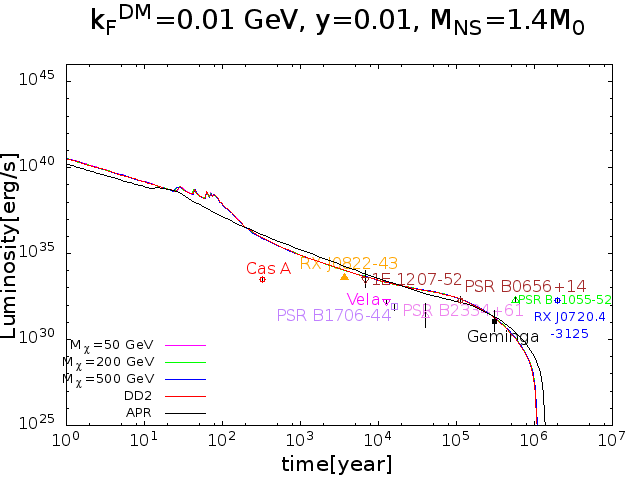}
\includegraphics[width=5.5cm,height=6.5cm]{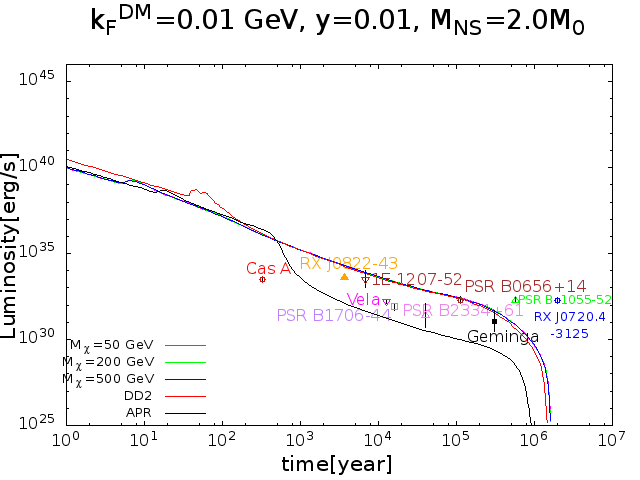}
\caption{Variation of luminosity with time for three differently chosen NS masses $M_{NS} = 1.0 M_{\odot}$  (Left panel), $M_{NS} = 1.4 M_{\odot}$ (middle panel), $M_{NS} = 2.0 M_{\odot}$ (right panel) 
with varying $M_\chi$ and fixed $k^{DM}_F=0.01$ GeV in each panel. The theoretical calculations are compared with the observational data of pulsars namely Cas A, RX J0822-43, 1E 1207-52, PSR B1706-44, Vela, PSR B2334+61, PSR B0656+14, Geminga, PSR B1055-52 and RX J0720.4-3125 shown by dots with error bars from left to right.}
\label{fig:11}
\end{figure}

\begin{figure}[H]
\includegraphics[width=5.5cm,height=6.5cm]{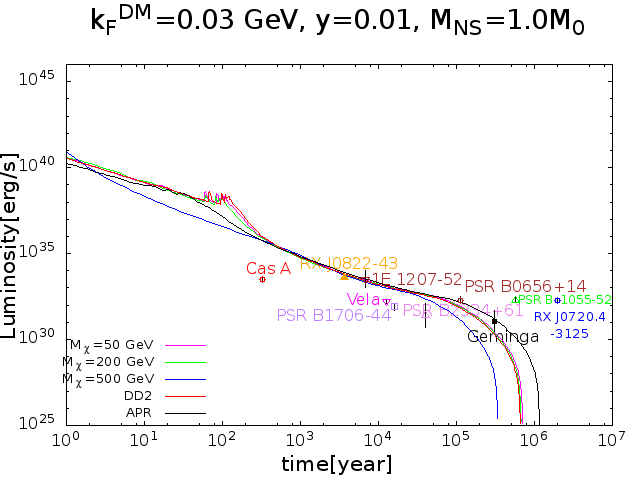}
\includegraphics[width=5.5cm,height=6.5cm]{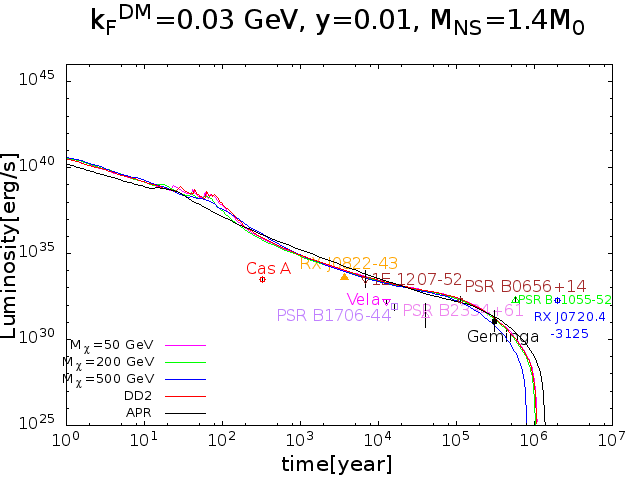}
\includegraphics[width=5.5cm,height=6.5cm]{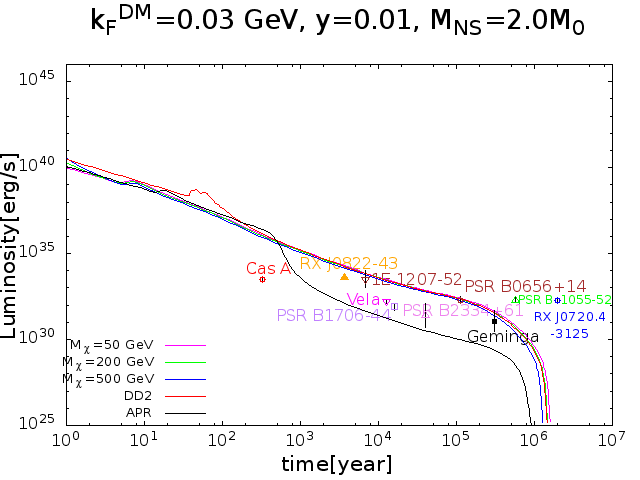}
\caption{Same as in Figure \ref{fig:11} but for $k^{DM}_F=0.03$ GeV}
\label{fig:12}
\end{figure}

\begin{figure}[H]
\includegraphics[width=5.5cm,height=6.5cm]{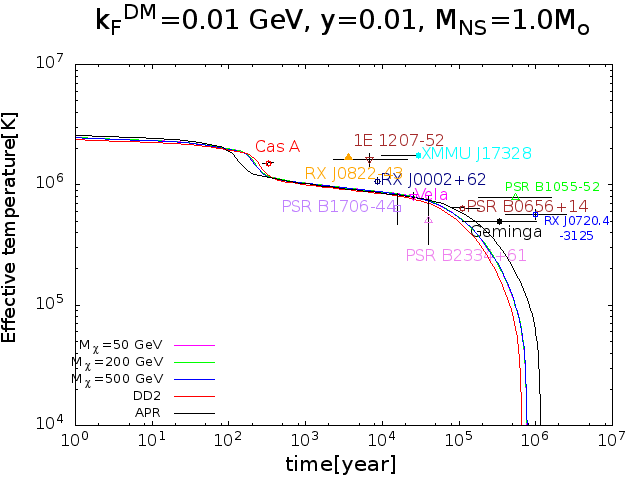}
\includegraphics[width=5.5cm,height=6.5cm]{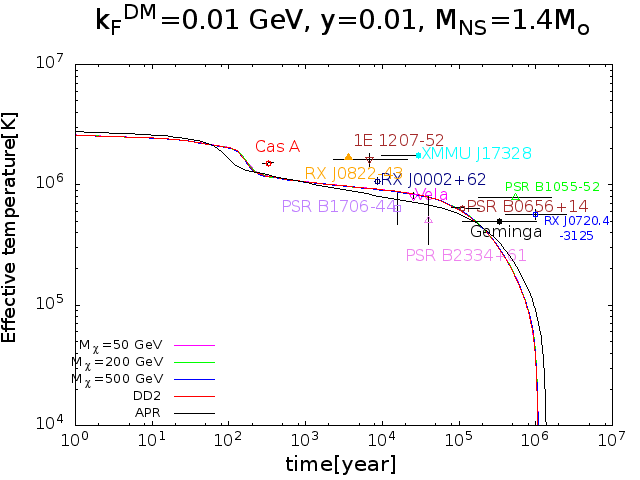}
\includegraphics[width=5.5cm,height=6.5cm]{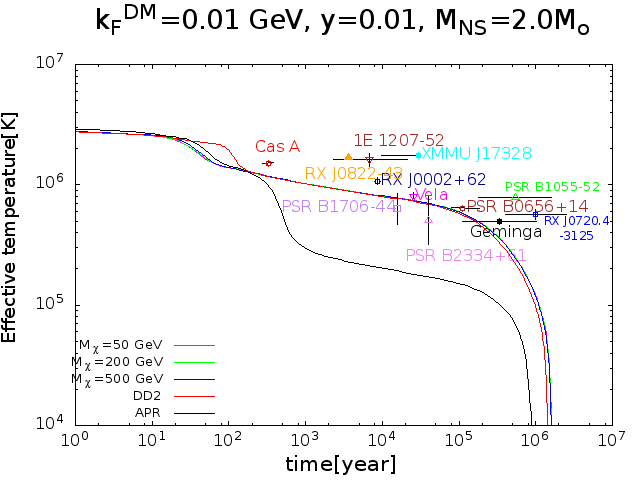}
\caption{Variation of tuminosity with time for chosen three different NS masses $M_{NS} = 1.0 M_{\odot}$  (Left panel), $M_{NS} = 1.4 M_{\odot}$ (middle panel), $M_{NS} = 2.0 M_{\odot}$ (right panel) 
with varying $M_\chi$ and fixed  $k^{DM}_F=0.01$ GeV in each panel. The theoretical calculations are compared with the observational data of pulsars namely Cas A, RX J0822-43, 1E 1207-52, RX J0002+62, XMMU J17328, PSR B1706-44, Vela, PSR B2334+61, PSR B0656+14, Geminga, PSR B1055-52 and RX J0720.4-3125 shown by dots with error bars from left to right.}
\label{fig:13}
\end{figure}

\begin{figure}[H]
\includegraphics[width=5.5cm,height=6.5cm]{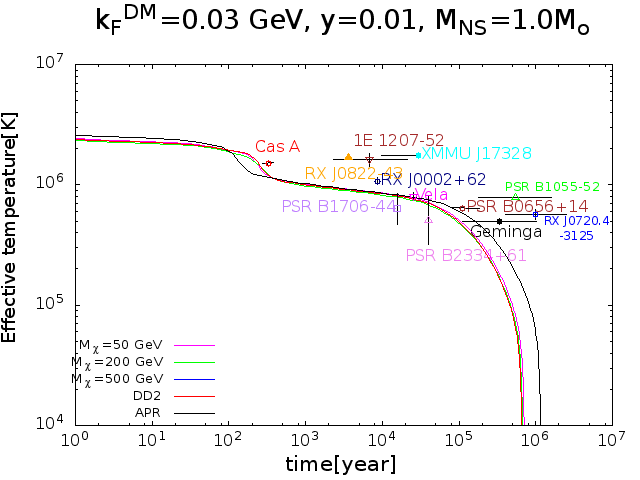}
\includegraphics[width=5.5cm,height=6.5cm]{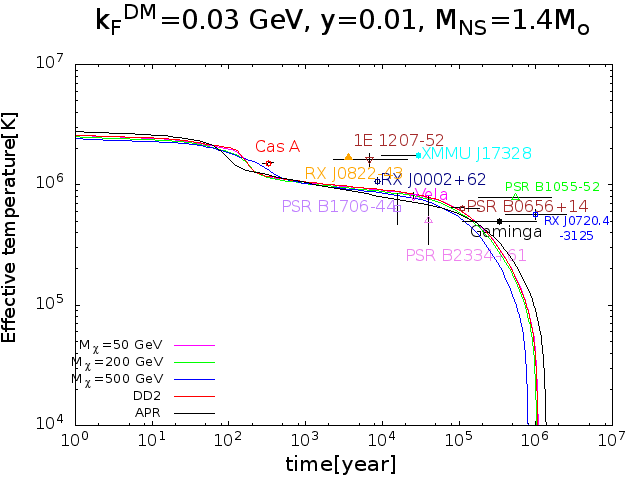}
\includegraphics[width=5.5cm,height=6.5cm]{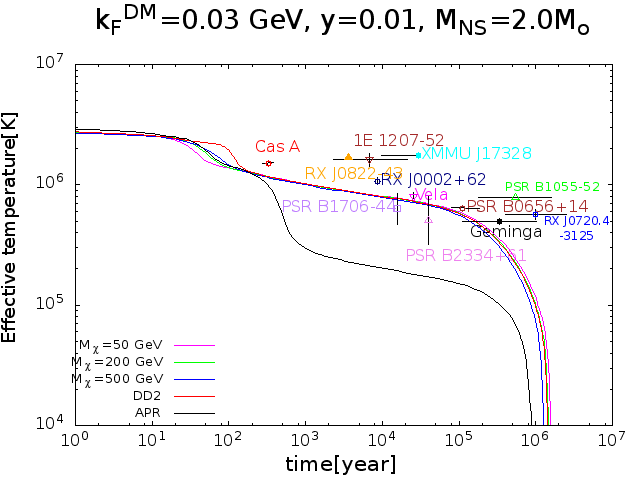}
\caption{Same as in Figure \ref{fig:13} but for$k^{DM}_F=0.03$ GeV}
\label{fig:14}
\end{figure}

\begin{figure}[H]
\includegraphics[width=8.5cm,height=8.5cm]{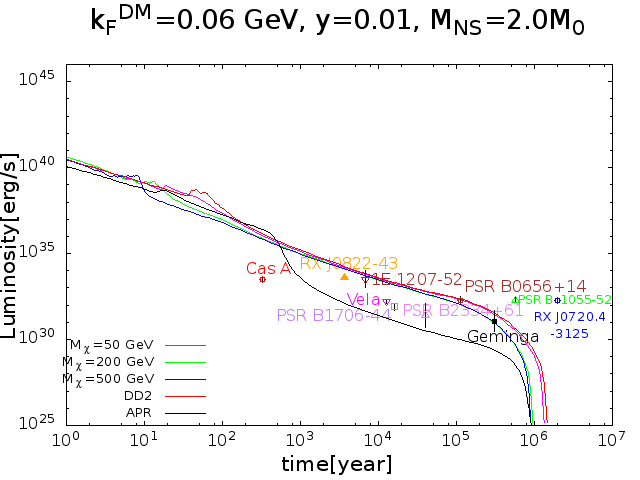}
\includegraphics[width=8.5cm,height=8.5cm]{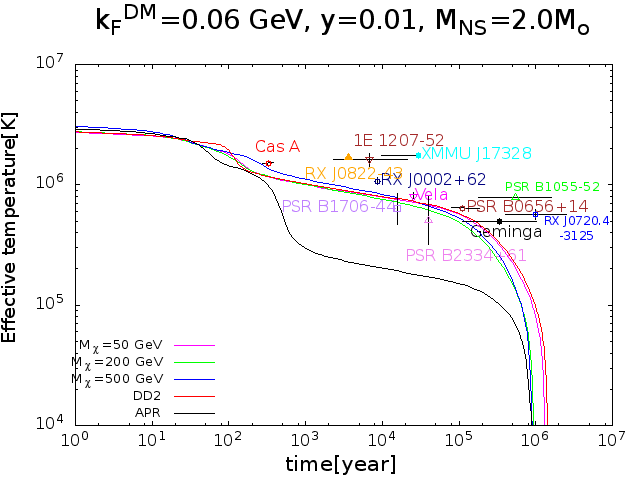}
\caption{Left panel is for luminosity vs time and right panel is for effective temperature vs time with varying $M_\chi$ and fixed  $k^{DM}_F=0.06$ GeV in both panels. The theoretical calculations
are also compared with the observational data of all chosen pulsars.}
\label{fig:15}
\end{figure}

We utilised NSCool Numerical code \cite{Nscool:2010dp} for studying cooling of NSs adopting different EOSs like APR, DD2 and DM admixed DD2. We considered different neutron star masses namely 1.0 $M_{\odot}$, 
1.4 $M_{\odot}$ and 2.0 $M_{\odot}$ for the calculations.
In case of DM admixed DD2, we explored the effect of variation of DM mass (50 GeV, 200 GeV and 500 GeV) and DM Fermi momentum $k^{DM}_F$ (0 GeV, 0.01 GeV, 0.02 GeV, 0.03 GeV and 0.06 GeV) 
on the cooling of NSs. It is important to mention here that $k^{DM}_F = 0$ GeV means dark matter density is zero but effective mass of nucleons will be effected due to non-zero Higgs-nucleon
Yukawa coupling (Eq. (\ref{eq:4})). For demonstrating DM-effect on neutron star cooling we plot the variations of luminosity with time  (Figures \ref{fig:5}-\ref{fig:7}) and effective temperature with time 
(Figures \ref{fig:8}-\ref{fig:10}).
As seen in all the plots (Figures \ref{fig:5}-\ref{fig:15}), shortly after birth,  
 NS cooling becomes dominated by  neutrino emitting processes as mentioned earlier. 
When the internal temperature has sufficiently dropped in nearly about $10^4-10^5$ year then
the cooling is dominated by photon emission from the NSs surface. 
In Figures \ref{fig:5}-\ref{fig:7}, luminosity vs time are plotted for different NS masses and for 
every NS mass diffferent EOSs are considered. For lower NS masses, cooling with DD2 is fastest and with APR it is slowest. Moreover, for fixed DM mass, cooling is faster for higher values
of DM Fermi momentum. But in heavier NSs, cooling with APR becomes fastest which might be due to appearance of direct Urca neutino emitting channels and variation due to $k^{DM}_F$ is the same as previously. The effect of $k^{DM}_F$ on cooling of heavier NSs becomes more and more prominent at higher values
of DM mass as is evident from
the right most panels of Figures \ref{fig:6} and \ref{fig:7}. Figures \ref{fig:8}-\ref{fig:10} are effective temperature vs time profiles and these can be explained the same way as Figures \ref{fig:5}-\ref{fig:7}. 
In Figures \ref{fig:11}, \ref{fig:12} 
and \ref{fig:15} (left panel) for luminosity vs time,
DM Fermi momentum is fixed and DM mass is varied for both heavier and lighter NSs. These figures clearly show  that the cooling is faster for higher values of DM mass and in this case also
cooling for heavier NSs is fastest with APR EOS. Figures \ref{fig:13}-\ref{fig:14} and \ref{fig:15} (right panel) are effective temperature vs time plots for lighter and heavier NSs where $k^{DM}_{F}$ is fixed and 
DM mass is varied. These
Figures can be explained the same way as Figures \ref{fig:11}, \ref{fig:12} and \ref{fig:15} (left panel).
In this case (Figures \ref{fig:11}-\ref{fig:14}), effect on cooling due to varying DM masses becomes more evident for lower mass NSs except for the case of $k^{DM}_F= 0.06$ GeV where variation due to
DM mass is prominent even for 
heavier mass NS (Figure \ref{fig:15}) which is due to very high DM Fermi momentum. For all masses of  DM admixed
NSs, all chosen DM Fermi momenta are considerably consistent with cooling data of pulsars namely PSR B0656+14, Geminga, Vela, PSR B1706-44 and PSR B2334+61 as evident from the Figures \ref{fig:5}-\ref{fig:15}. It is noted that Cas A barely agrees with the cooling curves corresponding to higher values of dark matter Fermi momenta and DM mass. These observed pulsars might contain dark matter (WIMP) with lower to moderate mass.
 Furthermore, as seen from left most panels of Figures \ref{fig:11}-\ref{fig:14}, it is evident that if
 small mass and super cold NSs are found in  future astronomical cooling observations we can say that heavier WIMPS may actually exist inside NSs.  



\begin{figure}[H]
\includegraphics[width=8.5cm,height=8.5cm]{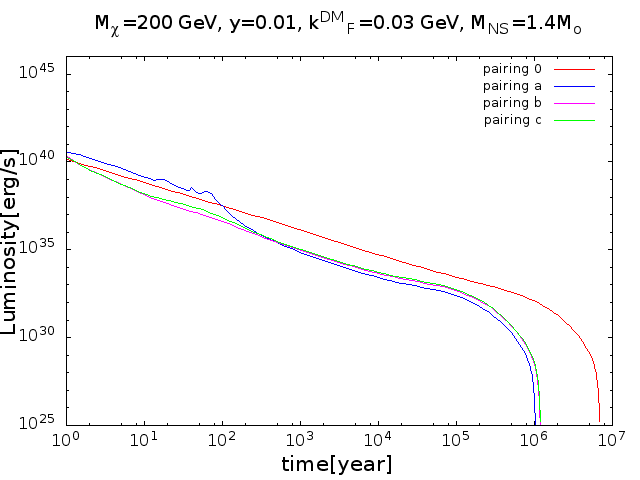}
\includegraphics[width=8.5cm,height=8.5cm]{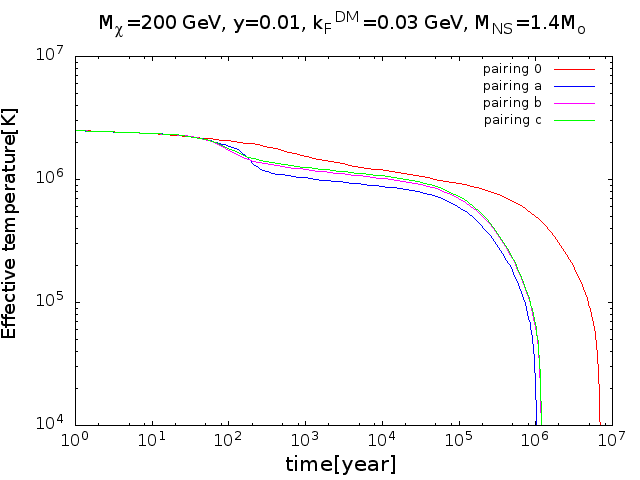}
\caption{Left panel is for luminosity vs time and right panel is for effective temperature vs time with different pairing gap models keeping $M_\chi$ and $k^{DM}_F$ fixed in both the panels.}
\label{fig:16}
\end{figure}

For demonstrating the effect of superfluidity on cooling, we plotted luminosity vs time and temperature vs time profiles in Figure \ref{fig:16} considering different pairing gap models inside the NSCool code because the actual value of  neutron \isotope[3]{P}$_2$
gap is unknown \cite{Nscoolguide}. In  Figure \ref{fig:16}, `pairing 0' means no pairing is considered and `pairing a ,b, c' correspond to three different \isotope[3]{P}$_2$
pairing gap  models. The details of these models are given in Ref. \cite{Nscoolguide} and the references therein.  It is evident that cooling is faster when pairing is considered and among the three pairing models, cooling with `pairing a' is slightly faster. Hence, we have considered `pairing a' model throughout the work.

\section{Summary and Conclusions}
In this work, we first prepared dark-matter admixed DD2 equation of state and explored the effect of dark matter mass and Fermi momentum on the neutron star equation of state. Dark matter-Higgs coupling is 
constrained using dark matter direct detection experiments namely XENON-1T, PandaX-II, LUX and  DarkSide-50. Then, we studied cooling of normal NSs using APR and DD2 equation of states and DM admixed neutron stars using 
dark-matter modified DD2 with varying dark matter mass and Fermi momentum for fixed DM-Higgs coupling. We have done our analysis by considering three neutron star masses one each from the lighter
(1.0 $M_{\odot}$), medium (1.4 $M_{\odot}$) 
and heavier (2.0 $M_{\odot}$) NSs. We demonstrate our results by choosing three different DM masses namely 50 GeV, 200 GeV and 500 GeV and different Fermi momenta $k^{DM}_F$ namely 0.01 GeV, 0.02 GeV, 0.03 GeV
and 0.06 GeV. We calculated the variations of luminosity and temperature of
the above mentioned neutron star masses with time and compared our calculations with the observed astronomical cooling data of pulsars namely Cas A, RX J0822-43, 1E 1207-52, RX J0002+62, XMMU J17328, PSR B1706-44, Vela, PSR B2334+61, PSR B0656+14, Geminga, PSR B1055-52 and RX J0720.4-3125. We found that APR EOS 
agrees well with the pulsar data for lighter and medium mass NSs whereas for DM admixed DD2 EOS, it is found for all considered NS masses, all chosen DM Fermi momenta are consistent with the observational data of PSR B0656+14, Geminga, Vela, PSR B1706-44 and PSR
B2334+61. Cooling becomes faster as compared to normal NSs in case of increasing
DM masses and  Fermi momenta. It is observed from the calculations that if low mass super cold NSs are observed in future that may support the fact that heavier WIMP can be present inside neutron stars. 

{}

\begin{thebibliography}{}
\bibitem{Glenda:1985} 
 N.K. Glendenning, Astrophys. J. {\bf 293}, 470 (1985).

\bibitem{Glenda:1997} 
 N.K. Glendenning, Compact Stars, Nuclear Physics, Particle Physics and General Relativity
(New York, Springer, 1997).

\bibitem{Rho:1979} 
 See Mesons in Nuclei, vol. 3, M. Rho, D. Wilkinson, Eds. (North-Holland, Amsterdam, 1979).

\bibitem{Kaplan:1986} 
 D.B. Kaplan, A.E. Nelson, Phys. Lett. B {\bf 175}, 57 (1986).

\bibitem{Collins:1975} 
 J.C. Collins, M.J. Perry, Phys. Rev. Lett. {\bf 34}, 1353 (1975).

\bibitem{Panotopoulos:2017idn} 
  G.~Panotopoulos and I.~Lopes,
  Phys.\ Rev.\ D {\bf 96}, 083004 (2017).

\bibitem{Das:2018frc} 
  A.~Das, T.~Malik and A.~C.~Nayak,
  Phys.\ Rev.\ D {\bf 99}, 043016 (2019).
  
\bibitem{Quddus:2019ghy} 
  A.~Quddus, G.~Panotopoulos, B.~Kumar, S.~Ahmad and S.~K.~Patra,
  arXiv:1902.00929 [nucl-th].
 
\bibitem{Bandyopadhyay:2018} 
D. Bandyopadhyay, S. A. Bhat, P. Char and D. Chatterjee Eur. Phys. J. {\bf A54}, 26 (2018), [arXiv: arXiv:1712.01715 [astro-ph.HE]].

\bibitem{Walecka:1974qa} 
  J.~D.~Walecka,
  Annals Phys.\  {\bf 83}, 491 (1974).
  
\bibitem{Serot:1997xg} 
  B.~D.~Serot and J.~D.~Walecka,
  Int.\ J.\ Mod.\ Phys.\ E {\bf 6}, 515 (1997).
 
 
\bibitem{Abbott:2017a} 
Abbott B P et al. Phys. Rev. Lett.{\bf 119}, 161101 (2017).
 
 
\bibitem{Abbott:2017} 
 Abbott B P et al. Astrophys. J. Lett. {\bf 848}, L13 (2017).
 
 
\bibitem{Cromartie:2019kug} 
  H.~T.~Cromartie {\it et al.},
  arXiv:1904.06759 [astro-ph.HE].
  

\bibitem{Weniger:2012tx} 
  C.~Weniger,
  JCAP {\bf 1208}, 007 (2012).

\bibitem{Bringmann:2012vr} 
  T.~Bringmann, X.~Huang, A.~Ibarra, S.~Vogl and C.~Weniger,
  JCAP {\bf 1207}, 054 (2012).
   
\bibitem{Adriani:2008zr} 
  O.~Adriani {\it et al.} [PAMELA Collaboration],
  Nature {\bf 458}, 607 (2009).
 
\bibitem{AMS-02}
M. Aguilar et al. [AMS collaboration], 
Phys. Rev. Lett. {\bf 113}, 121102 (2014).

\bibitem{Ambrosi:2017wek} 
  G.~Ambrosi {\it et al.} [DAMPE Collaboration],
  Nature {\bf 552}, 63 (2017).
  
  
  
  
\bibitem{Jungman:1995df} 
  G.~Jungman, M.~Kamionkowski and K.~Griest,
  Phys.\ Rept.\  {\bf 267}, 195 (1996).
  [hep-ph/9506380].

\bibitem{Paul:2018njd} 
  A.~Paul, D.~Majumdar and A.~Dutta Banik,
  JCAP {\bf 05}, 029 (2019), [arXiv:1812.10791 [hep-ph]].
  
\bibitem{Banik:2015aya} 
  A.~Dutta Banik, D.~Majumdar and A.~Biswas,
  Eur.\ Phys.\ J.\ C {\bf 76}, 346 (2016).
  
  
\bibitem{Biswas:2015sva} 
  A.~Biswas, D.~Majumdar and P.~Roy,
  JHEP {\bf 1504}, 065 (2015).
  
  
\bibitem{Biswas:2013nn} 
  A.~Biswas, D.~Majumdar, A.~Sil and P.~Bhattacharjee,
  JCAP {\bf 1312}, 049 (2013).
  
\bibitem{Peccei:2006as} 
  R.~D.~Peccei,
  Lect.\ Notes Phys.\  {\bf 741}, 3 (2008).
  
\bibitem{Weinberg:1977ma} 
  S.~Weinberg,
  Phys.\ Rev.\ Lett.\  {\bf 40}, 223 (1978).

\bibitem{Yaguna:2011qn} 
  C.~E.~Yaguna,
  JHEP {\bf 1108}, 060 (2011).

\bibitem{Molinaro:2014lfa} 
  E.~Molinaro, C.~E.~Yaguna and O.~Zapata,
  JCAP {\bf 1407}, 015 (2014).

\bibitem{Lidz:2018fqo} 
  A.~Lidz and L.~Hui,
  Phys.\ Rev.\ D {\bf 98}, 023011 (2018).

\bibitem{Amorisco:2018dcn} 
  N.~C.~Amorisco and A.~Loeb,
  arXiv:1808.00464 [astro-ph.GA].

  
\bibitem{Bergstrom:2006ny} 
  L.~Bergstrom, M.~Fairbairn and L.~Pieri,
  Phys.\ Rev.\ D {\bf 74}, 123515 (2006).
  
  
\bibitem{Fuller:2014rza} 
  J.~Fuller and C.~Ott,
  Mon.\ Not.\ Roy.\ Astron.\ Soc.\  {\bf 450}, L71 (2015).
  
   
  
\bibitem{Brayeur:2011yw} 
  L.~Brayeur and P.~Tinyakov,
  Phys.\ Rev.\ Lett.\  {\bf 109}, 061301 (2012).
  
\bibitem{Sandin:2008db} 
  F.~Sandin and P.~Ciarcelluti,
  Astropart.\ Phys.\  {\bf 32}, 278 (2009).
  
       
\bibitem{Page:2009fu} 
  D.~Page, J.~M.~Lattimer, M.~Prakash and A.~W.~Steiner,
  Astrophys.\ J.\  {\bf 707}, 1131 (2009).
  
\bibitem{Page:2004fy} 
  D.~Page, J.~M.~Lattimer, M.~Prakash and A.~W.~Steiner,
  Astrophys.\ J.\ Suppl.\  {\bf 155}, 623 (2004).

\bibitem{Yakovlev:2004iq} 
  D.~G.~Yakovlev and C.~J.~Pethick,
  Ann.\ Rev.\ Astron.\ Astrophys.\  {\bf 42}, 169 (2004).
  
\bibitem{Potekhin:2015qsa} 
  A.~Y.~Potekhin, J.~A.~Pons and D.~Page,
  Space Sci.\ Rev.\  {\bf 191}, no. 1-4, 239 (2015).
  
  
\bibitem{Potekhin:2017ufy} 
  A.~Y.~Potekhin and G.~Chabrier,
  Astron.\ Astrophys.\  {\bf 609}, A74 (2018).
 
\bibitem{Ding:2019}  
Wen-Bo Ding et al. Chinese Phys. Lett. {\bf 36}, 049701 (2019). 

\bibitem{Kouvaris:2007ay} 
  C.~Kouvaris,
  Phys.\ Rev.\ D {\bf 77}, 023006 (2008).
  
\bibitem{Hamaguchi:2019oev} 
  K.~Hamaguchi, N.~Nagata and K.~Yanagi,
  Phys.\ Lett.\ B {\bf 795}, 484 (2019).

 
\bibitem{Banik:2014} 
S. Banik, M. Hempel and D. Bandyopadhyay, Astrophys.J.Suppl. {\bf 214}, 22 (2014). 
   
\bibitem{Typel:2010} 
S. Typel et al., Phys. Rev. C {\bf 81}, 015803 (2010). 
 


\bibitem{Grigorian:2018bvg} 
  H.~Grigorian, D.~N.~Voskresensky and K.~A.~Maslov,
  Nucl.\ Phys.\ A {\bf 980}, 105 (2018).

\bibitem{Grigorian:2016leu} 
  H.~Grigorian, D.~N.~Voskresensky and D.~Blaschke,
  Eur.\ Phys.\ J.\ A {\bf 52}, 67 (2016).

\bibitem{Voskresensky:2001fd} 
  D.~N.~Voskresensky,
  Lect.\ Notes Phys.\  {\bf 578}, 467 (2001).
  
\bibitem{Yanagi:2019vrr} 
  K.~Yanagi, N.~Nagata and K.~Hamaguchi,
  Mon.\ Not.\ Roy.\ Astron.\ Soc.\  {\bf 492}, 5508 (2020).

\bibitem{Vigano:2013lea} 
  D.~Viganò, N.~Rea, J.~A.~Pons, R.~Perna, D.~N.~Aguilera and J.~A.~Miralles,
  Mon.\ Not.\ Roy.\ Astron.\ Soc.\  {\bf 434}, 123 (2013).


 
\bibitem{DeLuca:2004ck} 
  A.~De Luca, P.~A.~Caraveo, S.~Mereghetti, M.~Negroni and G.~F.~Bignami,
  Astrophys.\ J.\  {\bf 623}, 1051 (2005).
  

\bibitem{Sedrakian:2015krq} 
  A.~Sedrakian,
  Phys.\ Rev.\ D {\bf 93}, no. 6, 065044 (2016).

\bibitem{Umeda:1997da} 
  H.~Umeda, N.~Iwamoto, S.~Tsuruta, L.~Qin and K.~Nomoto,
  astro-ph/9806337.

  
\bibitem{Paul:2018msp} 
  A.~Paul, D.~Majumdar and K.~Prasad Modak,
  Pramana {\bf 92}, 44 (2019), [arXiv:1801.07928 [hep-ph]].
 

\bibitem{Akmal:1998cf} 
  A.~Akmal, V.~R.~Pandharipande and D.~G.~Ravenhall,
  Phys.\ Rev.\ C {\bf 58}, 1804 (1998).
   
\bibitem{Bhat:2018erd} 
  S.~A.~Bhat and D.~Bandyopadhyay,
  J.\ Phys.\ G {\bf 46} , 014003 (2019),
  [arXiv:1807.06437 [astro-ph.HE]].

\bibitem{Banik:2002} 
S. Banik and D. Bandyopadhyay, Phys. Rev. C {\bf 66}, 065801 (2002).

\bibitem{Typel:2005} 
S. Typel, Phys. Rev. C {\bf 71},  064301 (2005).

\bibitem{Typel:1999} 
S. Typel and H. H. Wolter, Nucl. Phys.A {\bf 656}, 331 (1999).


\bibitem{cline:2015} 
J. M. Cline, K. Kainulainen, P. Scott and C. Weniger, Phys. Rev. D 88, 055025 (2013), Erratum: Phys.
Rev. D {\bf 92}, 039906 (2015).


\bibitem{Li:2012qf} 
  X.~Li, F.~Wang and K.~S.~Cheng,
  JCAP {\bf 10}, 031 (2012).


\bibitem{Oppenheimer:1939} 
 J. R. Oppenheimer and G. M. Volkoff, Phys. Rev. {\bf 55}, 374 (1939).

\bibitem{Aprile:2015uzo} 
  E.~Aprile {\it et al.} [XENON Collaboration],
  JCAP {\bf 1604}, 027 (2016).
 
\bibitem{Tan:2016zwf} 
  A.~Tan {\it et al.} [PandaX-II Collaboration],
  Phys.\ Rev.\ Lett.\  {\bf 117}, 121303 (2016).
    
\bibitem{Akerib:2016vxi} 
  D.~S.~Akerib {\it et al.} [LUX Collaboration],
  Phys.\ Rev.\ Lett.\  {\bf 118}, 021303 (2017).
  
\bibitem{Marini:2016haq} 
  L.~Marini {\it et al.} [DarkSide Collaboration],
  Nuovo Cim.\ C {\bf 39}, 247 (2016).
  
\bibitem{Page:2005fq} 
  D.~Page, U.~Geppert and F.~Weber,
  Nucl.\ Phys.\ A {\bf 777}, 497 (2006).
 
\bibitem{Nscool:2010dpl} 
http://www.astroscu.unam.mx/neutrones/NSCool/. Dany Page,
Cooling of
Neutron Stars, Lecture 2.  

\bibitem{Nscool:2010dp} 
http://www.astroscu.unam.mx/neutrones/NSCool/. We use the
 file Crust-EOS-Cat-HZD-NV.dat for the equation of state input.
 
  
\bibitem{Yakovlev:2000jp} 
  D.~G.~Yakovlev, A.~D.~Kaminker, O.~Y.~Gnedin and P.~Haensel,
  Phys.\ Rept.\  {\bf 354}, 1 (2001).
    
  
  
 
  
\bibitem{Nscoolguide}   
http://www.astroscu.unam.mx/neutrones/NSCool/NSCool$\_$Guide$\_$0.0$\_$Control.pdf.
       
\end{thebibliography}
\end{document}